# Tackling COVID-19 through responsible AI innovation: Five steps in the right direction

**David Leslie**[*]
THE ALAN TURING INSTITUTE
96 Euston Road, London NW1 2DB
United Kingdom

**Abstract**

Innovations in data science and artificial intelligence/machine learning (AI/ML) have a central role to play in supporting global efforts to combat COVID-19. The versatility of AI/ML technologies enables scientists and technologists to address an impressively broad range of biomedical, epidemiological, and socio-economic challenges. This wide-reaching scientific capacity, however, also raises a diverse array of ethical challenges. The need for researchers to act quickly and globally in tackling SARS-CoV-2 demands unprecedented practices of open research and responsible data sharing at a time when innovation ecosystems are hobbled by proprietary protectionism, inequality, and a lack of public trust. Moreover, societally impactful interventions like digital contact tracing are raising fears of "surveillance creep" and are challenging widely-held commitments to privacy, autonomy, and civil liberties. Pre-pandemic concerns that data-driven innovations may function to reinforce entrenched dynamics of societal inequity have likewise intensified given the disparate impact of the virus on vulnerable social groups and the life-and-death consequences of biased and discriminatory public health outcomes. To address these concerns, I offer five steps that need to be taken to encourage responsible research and innovation. These provide a practice-based path to responsible AI design and discovery centered on open, accountable, equitable, and democratically governed processes and products. When taken from the start, these steps will not only enhance the capacity of innovators to tackle COVID-19 responsibly, they will, more broadly, help to better equip the data science and AI/ML community to cope with future pandemics and to support a more humane, rational, and just society.

## 1 Introduction

In June 1955, the great Hungarian mathematician and polymath John Von Neumann published a popular essay entitled, "Can we survive technology?" (Von Neumann, 1955). Von Neumann, then stricken with terminal cancer, wrote about what he called "the maturing crisis of technology," a situation in which the global effects of accelerating technological advancement were outpacing the development of ethical and political self-understandings that were capable of responsibly managing such an explosion of innovation. This crisis, he feared, was creating unprecedented dangers of species-level self-destruction ranging from geoengineering and unbridled automation to nuclear holocaust. At the same, he puzzled that "technological power, technological efficiency as such" was "an ambivalent achievement." That is, the very forces of ingenuity that were creating the dangers of anthropogenic self-annihilation contained within themselves the potential to benefit humanity. They possessed, *in potentia*, a kind of countervailing redeeming power.

As society now grapples with a different kind of crisis than the one Von Neumann had in mind, his reflections are no less relevant for thinking through data-driven technology's direction of travel in confronting the challenges presently faced. The maturing crisis of technology to which he referred applies especially to the field of artificial intelligence (AI) and machine learning (ML).[1] In less than a generation, exponential leaps in information processing power have coalesced with the omnipresent data extraction capabilities of an ever more dynamic, integrated, and connected digital world to provide a fecund spawning ground for the explosion of AI/ML technologies. And, as these innovations have advanced apace—as the scope of their impacts has come to stretch from the most intimate depths of self-development to the fate of the biosphere itself—we need ever more to reflect soberly on Von Neumann's worry: Have we developed the novel ethical and political self-understandings, values, practices, and forms of life necessary to responsibly steer and constrain the rapid proliferation of AI/ML technologies?

[*] Correspondence concerning this article may be sent to the author by email: dleslie@turing.ac.uk

By all accounts, society has, in fact, struggled to keep up. The all-too-common "break first, think later" attitude of many of those at the wheel of commercial AI/ML innovation has been a recipe for financial success simultaneously as it has been fast track in the race to the ethical bottom. Prominent examples of algorithmic bias and discrimination, of proprietary black boxes and organisational opacity, and of macroscale behavioural tracking, curating, and nudging have led to social consternation and distrust. More troubling still, AI/ML-enabled capabilities for hyper-personalized targeting, anticipatory calculation, and algorithmic administration at scale have manifested in intrusive hazard-pre-emption regimes (O'Grady, 2015) ranging from data-driven border control (Amoore, 2009; Amoore and Raley, 2017) to predictive policing and commercial surveillance. They have also enabled emergent digital autocracies to engage in population-level individual monitoring and mass disciplinary control. Most consequentially, though, global prospects for a divisive geopolitical sprint to technological ascendency in AI/ML are now opening up new possibilities for destructive struggles of an unprecedented scale. In virtue of the accelerating pace of digital innovation propelled by the hasty pursuit of competitive advantage, such conflicts may soon pose very real dangers to the future of life itself—dangers extending from calamitous cyberattacks on infrastructural vulnerabilities, algorithmically streamlined biological warfare, and human enhancement "arms races" to smart nuclear terrorism and the potentially genocidal proliferation of lethal autonomous weapons.

All this would seem to leave us, then, at the perilous crossroads of two crises—one rooted in the destructive potentials of our extant technological practices and another demanding that those same practices be marshalled as a saving power to combat the destruction inflicted by an inhuman biological agent. Faced with the current public health crisis, data scientists and AI/ML innovators may be inclined to ask: Are we ready for this? Can we find a responsible path to wielding our technological efficacy ethically and safely? In what follows, I claim that this crossroads need not induce paralysis as to which way we should go, and, in fact, presents us with clear signage for finding the right way forward. When pressed into the service of the public good, biomedical AI/ML applications have already made noteworthy progress in assisting doctors and researchers in the areas of diagnostics, prognostics, genomics, drug discovery, epidemiology, and mobile health monitoring.[2] And, though all of these areas of advancement hold the great promise of helping healthcare professionals to combat COVID-19, they also come with substantial ethical hazards. What we need now are actionable means to navigate these.

---

[1] In the following, I will use the abbreviation AI/ML to indicate those information processing systems or algorithmic models that intervene in the human world (directly or through the insights they enable) by carrying out cognitive or perceptual functions previously reserved for human beings. This broadly pragmatic and functionalist definition is meant to be as generally applicable to both deterministic and non-deterministic algorithm-based computing machinery as it is non-metaphysical (IEEE-USA, 2017; Leslie 2019b; Minsky, 1968; OECD, 2019a, 2019b). Similarly, I will refer to "data science" as the broad, interdisciplinary set of approaches and techniques that combine statistics, applied mathematics, data mining, computer programming, and other related fields to gain a better conceptual understanding and practical grasp of the data patterns underlying the empirical world.

[2] Some examples of these contributions include: In diagnostics, detecting diabetic retinopathy (Gulshan et al., 2016), using waveform analysis to identify birthing paths (Fergus et al., 2017), early diagnosis of Alzheimer disease (Liu et al., 2014), detecting lymph node metastases in women with breast cancer (Bejnordi et al., 2017), classifying sepsis in emergency departments (Horng, 2017), using clinical measurements to classify patients in a pediatric ICU (Lipton et al., 2015), classifying skin cancer (Esteva et al., 2017), diagnosing acute coronary syndrome (Berikol, 2016); In prognostics, predicting breast cancer survival (Katzman et al., 2018), predicting heart condition

Here, I argue that, in our current dilemma, the data science and AI community, writ large, ought to draw upon the hard-gained critical leverage and normative resources provided by applied ethics, responsible research and innovation (RRI), science and technology studies (STS), and AI/ML ethics to close the gap between fleetfooted innovation and slow-moving ethical and social values. In the first two sections I will motivate this deliberate turn to responsible AI innovation. Starting with the unparalleled challenges presented to data scientists and AI/ML researchers by the pandemic, I will explore how some of the more intractable ethical issues already faced by AI/ML innovation are raising their heads in the present circumstance of a global public health crisis that is placing researchers under unprecedented pressures to rapidly respond. I will then lay out some of the ethical pitfalls and societal challenges faced by the potential introduction of digital contact tracing and health monitoring technologies into a networked, big data society that is in peril of rocking back and forth between the Scylla of mass digital surveillance and the Charybdis of an ethically chilling but privacy-securing automation-all-the-way-down. Finally, I will move on to offering five steps toward responsible AI/ML research and innovation that need to be taken to address these concerns. Drawing upon current thinking in applied AI/ML ethics, social scientific approaches to data-driven technologies, and RRI, these steps suggest a means of attaining and assessing open, accountable, equitable, and democratically governed AI/ML processes and products.

## 2 Combatting COVID-19 on the second front: New challenges, old problems

The tasks that face us as a society, at present, are posing extraordinary ethical challenges of a kind that many of us have never before experienced. On the frontlines of the pandemic, our healthcare professionals are confronted with a merciless convergence of limited resources and surging illness. They are having to make difficult life-and-death choices about who receives critically needed care and how. As they do battle in the global struggle against the pandemic, these heroes, and all the essential workers who keep the ship of society afloat in times of crisis, face gruelling and unprecedented demands for self-sacrifice, moral fortitude, and resilience. These trials of conscience and character are testing the depths and shallows of us all and transforming our lives forever.

But the global struggle against COVID-19 is also being fought on a second crucial front with its own set of broad-reaching ethical challenges. While so many of our doctors, nurses, and key workers combat the virus on the frontlines, researchers and technologists too must tirelessly labour on the frontiers of biomedical, epidemiological, and societal innovation, so that their scientific discoveries can be employed to manage the spread of the virus and mitigate its effects.

In the data science and artificial intelligence (AI) community, such second-front efforts are already well under way. Machine learning (ML) and data-driven technologies are already augmenting human capacities to better tackle the challenges of the pandemic (Bullock et al., 2020). These AI/ML-assisted interventions range from AI-supported radiological diagnostics (Ai et al., 2020; Gozes et al., 2020; Shan et al., 2020; Wang et al., 2020;), prognostics based on clinical data (Pourhomayoun & Shakibi, 2020; Yan et al., 2020;

related hospitalization (Brisimi, 2018), predicting outcomes in colorectal cancer (Bychkov, 2018), predicting outcomes in non-small cell lung cancer (Yu et al., 2016); In genomics, predicting the sequence specificities of DNA- and RNA-binding proteins (Alipanahi et al., 2015), denoising genome-wide histone ChIP-seq (Koh et al., 2016), predicting protein structures from protein sequences (Lyons et al., 2014); In epidemiology, understanding outcomes in community-spread pneumonia (Wiemken, 2020), understanding degenerative diseases (Nathanson, 2019), detecting food-born illness (Sadilek et al., 2018); In mobile monitoring, diagnosing heart failure through wearable technology monitoring (Inan et al., 2018), estimating energy expenditure with wearable sensors (Zhu et al., 2015). For good additional landscape views, see: (Miotto et al., 2017; Panch et al., 2018; Stephenson, 2019; Wainberg et al., 2018).

Qi et al., 2020), pharmaceutical discovery and repurposing (Beck et al., 2020; Hu, Jiang, & Yin, 2020), and test-kit development (Metsky et al., 2020) to methods of protein and RNA profiling that are shedding light on virus function and disease progression (Jumper et al., 2020; Senior et al., 2020; Zhang et al., 2020). Likewise, in the area of research support, vast troves of existing biomedical literature are being mined by AI/ML technologies to identify clinically established drugs and treatment methods that may be of use in fighting the SARS-CoV-2 infection (Ge et al., 2020). Taken cumulatively, such interventions are helping to greatly enhance the quality and speed of the human response to the outbreak.

In the areas of epidemiological modelling and social-demographic analysis too, the high-dimensional processing capacity of AI/ML applications are helping scientists to generate more effective real-time forecasts of the spread of infection and of the locations of potential outbreaks (Al-qaness et al., 2020; Hu et al., 2020). Signally, at the very outset of the pandemic an AI/ML system from the Canadian health monitoring platform, BlueDot, warned of the outbreak nearly two weeks before the World Health Organisation made its own announcement (Marks, 2020; Niiler, 2020). AI/ML-supported population-level insight is also being used to combat the dissemination of misinformation about the pandemic (the spread of the so-called "infodemic") (Boberg et al., 2020; Chen, Lerman, & Ferrara, 2020; Cinelli et al., 2020; Mejova & Kalimeri, 2020). Better knowledge about the reach and sources of the propagation of misinformation will help to produce more effectual policy-interventions and to promote more critical information consumption at scale.

From a wide-angled view, this expansive spectrum of AI-supported interventions demonstrates an unprecedented opportunity for the data science and AI community to press its energies and talents into the service of advancing the public good. And yet, the novel practical and sociotechnical challenges posed by the current coronavirus pandemic suggest that members of this broad church must proceed with a heightened sobriety and vigilance. The prevalent urgency for answers and the pains and pressures of producing highly impactful research in the context of a global health crisis are only magnifying many of the existing ethical concerns raised by the use of AI/ML in medicine, epidemiology, and public health, even in normal times.

This has already been well-illustrated in a timely review of 31 prediction models from 27 early studies of COVID-19 by Wynants et al. (2020). In their critical appraisal, the authors find these models to be "at high risk of [statistical] bias, mostly because of non-representative selection of control patients, exclusion of patients who had not experienced the event of interest by the end of the study, and high risk of model overfitting" (p. 1). This risk of bias is attributed to "poor reporting and poor methodological conduct for participation selection, predictor description, and statistical methods used" (p. 7). The review also highlights the fact that the 12 diagnostic imaging studies of CT scans at hand lacked clear information on how the data was pre-processed and presented highly complex algorithms that transformed imaging data into predictors in opaque and unintelligible ways. Though the authors acknowledge that these studies, as a whole, were "done under severe time constraints caused by urgency," they also caution that, in a highly distressed clinical environment, practitioners might be encouraged to "implement prediction models without sufficient documentation and validation," leading to potentially harmful outcomes (pp. 8-9). Each of the issues raised by Wynants et al. is worthy of some unpacking.

## 2.1 Pitfalls of COVID-19-related research I: Algorithmic bias and discrimination

First, complaints about selection biases and the representativeness of the datasets used to build the diagnostic, prognostic, and resource-management-level prediction models in question tap into deeper concerns about how mismatches between data samples and target populations can lead to deleterious or discriminatory outcomes. It has long been recognized that insufficient cohort diversity and the under- or over-representativeness of datasets can lead AI/ML systems trained on this data to have biased and inequitable impacts on certain subpopulations (Barocas and Selbst, 2016; Calders & Zioblaite, 2013; Ferryman & Pitcan, 2018; Lehr & Ohm, 2017; O'Neill, 2016). Corresponding equity issues in data-driven approaches to medicine can, for example, arise in electronic health records (EHR) that fail sufficiently to include members of disadvantaged or marginalized groups who are unable to access the healthcare system (Arpey et al., 2017; Gianfrancesco et al., 2018) or in the sample selection biases that emerge when data availability is limited to well-resourced, digitally mature hospitals that disproportionately serve a particular racial or socioeconomic segment of a population to the exclusion of others.

Beyond dataset inequities, healthcare relevant patterns of discrimination and bias arise throughout the AI/ML production workflow, from biased choices made in data pre-processing and feature engineering to the ways in which various model parameters are tuned over the course of model design and testing (Berk et al., 2017; d'Alessandro et al., 2017; Kamiran & Calders, 2012; Leslie, 2019a; Suresh & Guttag, 2017). Of particular concern are the potentials for discriminatory harm that surface at the level of problem formulation, namely, in the ways that data scientists and AI innovators define target variables and identify their measurable proxies (Passi & Barocas, 2019). Definition-setting determinations made by design teams and researchers can perpetuate and reinforce structural inequalities and structural injustices (Jugov &Ypi, 2019; Young, 1990, 2009, 2011) by virtue of biased assumptions that creep into the way solutions are devised and measurements moulded. This form of discrimination can have an especially devastating effect in the field of health policy, as Obermeyer et al. (2019) demonstrated in their examination of how the label choice made for a commercial risk prediction tool in US healthcare led to systemic discrimination against millions of black patients, who tended to be far sicker than whites at an equivalent risk score.[3]

If left unattended to—especially in view of the current design-time pressures placed on project teams for rapid responses and insights—these sociotechnical tendrils of algorithmic bias and discrimination may only further tighten their grip on AI-supported practices and outcomes. This ethical hazard is, in fact, made worse by the disproportionately harmful effects of the COVID-19 pandemic on disadvantaged and vulnerable communities that are already subject to significant health inequities as well as overly susceptible to catastrophic, disaster-related harms (Bolin & Kurtz, 2018; Fothergill & Peek, 2004; Kleinenberg, 2015; Kristal et al., 2018; Van Bavel et al., 2020;Wang & Tang 2020). Indeed, such coronavirus-linked harmful effects are creating a kind of vicious discriminatory double punch whereby existing biases that make inroads into the healthcare-related algorithmic tools and applications created to combat the illness may harm disadvantaged people more because of the high, safety-critical impact these technologies have on them simultaneously as these biases may harm more disadvantaged people due to the disproportionate damage being inflicted on them by the virus itself.

## 2.2 Pitfalls of COVID-19-related research II: Adverse data impact

A second issue raised by Wynants et al. has to do with data impact, *viz.* the need, in diagnostic, prognostic, and policy-level prediction models, for complete, consistent, accurately measured, relevant, and timely data that is of sufficient quantity to produce reliable out-of-sample generalization (Ehsani-Moghaddam, 2019; Hienrich et al., 2007; Wang et al., 1995). Wynants et al. highlight the fact that the studies that they appraise face the common hazard that small sample sizes (drawn from scarce and geographically limited patient populations) will lead to overfitting and compromised generalizability (Foster et al., 2014; Riley et al., 2020; Wynants et al., 2020). They note that "immediate sharing of well documented individual participant data from COVID-19 studies is needed for collaborative efforts to develop more rigorous prediction models and validate existing ones" (p. 7).

[3] The problem here was that the designers of the model chose health care costs as the measurable proxy for the target concept of ill health. That the former is an insufficient stand-in for the latter becomes clearer when one considers factors such as (1) the challenges to accessing healthcare faced by traditionally disadvantaged subpopulations and (2) the challenge of the reduced trust in medical services experienced by historically maltreated social groups. This directly affects their level of engagement in healthcare systems. Because of correlations between socioeconomic status and race, black patients (even the insured) are less likely to run up the same level of healthcare costs as whites of greater advantage. These insights about erroneous proxies in (Obermeyer et al., 2019) follow on from the earlier work of two of the authors on mismeasurement in heath policy applications (Mullainathan & Obermeyer, 2017).

Extant issues with data quality and responsible data sharing in the healthcare domain will likely pose challenges here of which AI/ML researchers and innovators should take heed. Hindrances to the access and availability of sufficiently high-quality data in the context of a global public health emergency present difficulties that augment the effects of the widespread tendencies to health data silo-ing that have generated a motley of non-integrated data formats (Shortliffe & Sepúlveda, 2018) and a wide variability in data quality and integrity (He et al., 2019; Hersh et al., 2013; Kruse et al., 2016; Verheij et al., 2018). Prevailing gaps in digital maturity across hospitals, regions, and countries may also act as roadblocks to accessing data of sufficient quality and quantity to pick up generalizable and transportable signals from target populations. The general lack of preparedness to mobilize digital information on the second front has already been evidenced in the scramble to rapidly hand-code COVID-19 symptom checker chatbots in lieu of training data accurate enough to pursue more sophisticated AI/ML methods (Kohler & Scharte, 2020).

Other data-mobilizing suggestions that have been made by AI/ML researchers confronting SARS-CoV-2-related clinical questions raise an equally vexing set of data quality and sharing concerns. Van der Schaar and Humphries (2020) have proposed to link EHRs with passive data from pervasive sensing and mobile technologies in order "to issue accurate predictions of risk and help uncover the social structures through which systemic risks manifest and spread" (van der Schaar and Humphrey, 2020, p. 2). While forward-looking, these proposals face obstacles in terms of dataset representativeness and well-established uncertainties in the quality of unstructured big data (Bailly et al., 2017; Cahan et al., 2019; Kruse et al., 2016; Miotto et al., 2017). They also introduce long-concerning ethical risks related to informational privacy, de-identification, and informed consent—specifically, as these principles relate to the collection, linking, and use of passive data exhaust containing sensitive and potentially de-anonymizing information (De Montjoye, 2015; Golle, 2006; Klasnja et al., 2009; Maher et al., 2019; Ohm, 2010; Smith et al. 2016; Sweeney, 2001).

## 2.3 Pitfalls of COVID-19-related research III: Lack of process transparency and model interpretability

A final set of issues raised by Wynants et al. can be grouped, by family resemblance, into the category of transparency. In the review, the authors emphasize the prevalence in the appraised studies both of the low quality and opaqueness of research methods and recording practices and of the opaqueness and lack of interpretability of the predictive models themselves. The first of these problems can be classified as insufficient process transparency, the second, insufficient outcome transparency (Leslie, 2019a). To take the former first, having transparent organizational and research practices as well as well-reported documentation of them becomes all-the-more vital in the context of the COVID-19 pandemic inasmuch as normal protocols that govern patient consent and privacy may be suspended, amended, or compromised. The absence of explicit sanction places a higher burden of transparency and accountability on researchers, who must ensure that their research practices are worthy of justified public confidence and trust. Even in recent, non-crisis times, concerns about a lack of this kind of process transparency, at the levels of both organisational conduct and research practice (van der Aalst et al., 2017), have prompted demands for better approaches to operationalizing answerability and auditability in healthcare-related AI/ML innovation, so that the public can be reassured that their health data are safely and responsibly being used for advancing patient and community wellbeing (Habli et al., 2020; Hays et al., 2015; Spencer et al., 2016; Stockdale et. al., 2019).

At a more basic level, anxieties about process transparency have already had a direct bearing on the adoption, application, and effectiveness of data-driven decision support in a broad range of clinical environments. Poor methodological conduct and reporting, across many different areas of research in clinical prediction modelling (Collins et al., 2013; Damen et al. 2016; Mallet et al., 2010; Siontis et al., 2015), has led to limitations in the perceived reliability and applicability of such studies in decision support settings (Bouwmeester, 2012). Poor reporting and unclear methodological conduct function as a stumbling block for the reproducibility of results, and these are then often met with justified trepidations by clinicians. Without externally validated research that may be corroborated through replicated experimental methods and that supports generalizability to out-of-sample instances (Altman et al. 2009; Moons et al., 2012), clinical uptake will be significantly limited (Collins, 2014; Khalifa et al., 2019; Vollmer et al., 2019).

Similar degrees of reasonable mistrust have been generated among clinicians and patients due to a lack of transparency in the innerworkings and underlying rationale of the decision-assistance models themselves. As supports for evidence-based reasoning in medical and public health practices, diagnostic, prognostic, and policy-level prediction models bear the burden of having to be optimally intelligible, understandable, and accessible to clinical users and affected individuals (ARMC, 2019; Doshi-Velez & Kim, 2017; Gilvary, 2019; ICO, 2019; Jia et al., 2020; Miller, 2018; Nauck & Kruse, 1999; Rudin, 2019; Vellido et al, 2012; Vellido 2019; Wainberg et al., 2018). The optimization of model interpretability enables data scientists to build explanatory bridges to clinicians (Lakkaraju, 2016), who can then draw upon a given model's processing results to justify evidence-driven clinical decision-making (Shortliffe & Sepúlveda, 2018; Tonakeboni et al., 2019). Such bridges allow for gains in the objectivity and robustness of clinical judgment by making possible the detection of a greater range of patterns drawn from the vast complexity of underlying data distributions accessible to practitioners (Morley et al., 2019). Moreover, a high degree of interpretability allows data scientists and end users to better understand why things go wrong with a model when they do; as such it can help them to continually evaluate a model's limitations while scoping future improvements. Bridges between data science and clinical practice also allow clinicians to make sense of (and hence, to make better use of) inferences and insights derived from trained models in the contexts of their application domains (Holzinger et al., 2017).[4] Some have even claimed that medical ethics require interpretable AI systems insofar as the doctors who use them to support their care of patients must be able to provide meaningful information about the logic behind the treatments chosen and applied (Vayena et al., 2018).

Regardless of the wide acceptance of these desiderata of AI/ML interpretability in medicine and public health, an unresolved tension remains within the concept of outcome transparency—one with significant ramifications for an innovation environment influenced by the exigencies of the pandemic response. This has to do with the oft discussed trade-off between performance (predictive accuracy or other metrics) and interpretability (Ahmad et al., 2018; Bologna and Hayashi, 2017; Breiman, 2001; Caruana, 2015; Freitas, 2013; Gunning, 2017; He et al., 2019; Holzinger et al., 2019; Selbst & Barocas, 2018). The conventional view suggests that the deployment of complex AI/ML model classes (like deep learning or ensemble methods) leads, in general, to a boost in model performance in comparison to simpler techniques (such as regression- or rule-based methods), but only at the expense of interpretability. Thinking in the context of high stakes decision-making, Cynthia Rudin has recently characterized the idea that this trade-off is necessary as a "myth" (Rudin, 2019, p. 2). She argues that in domains like medicine—where much of the clinical data are organically representable as meaningful features and well-structured—interpretable algorithms have roughly equivalent performance as more opaque techniques. At the same time, the native understandability of such algorithms eliminates the need for surrogate, post-hoc explanatory mechanisms that tend to have low-fidelity to the slippery non-linearity of their black box counterparts.[5]

[4] This domain-sensitivity is crucial in medical decision support systems: Only when interpretable models are designed with a proper understanding of the missing data mechanisms endemic to the messiness of the clinical environment of concern, can they generate outputs that are appropriately responsive to its complexities and uncertainties (Ghassemi, 2018).

Rudin's steer is away from an explainability culture (that begins by reaching for the black box and then tries to find simpler auxiliary models to elucidate it) and towards an interpretability culture that starts with attempts to produce interpretable models through solid knowledge discovery and careful model iteration.[6] This prioritization of outcome transparency is consistent with positions held by clinicians who see AI/ML decision-support systems as bolstering evidence-based medical practice by widening and enriching the informational background for the exercise of human judgment (Shortliffe & Sepúlveda, 2018; Tonekaboni et al., 2019). However, others have pointed out that the utility of high performance, high interpretability models in clinical environments has yet to be demonstrated due to the infrequency of their application (Ahmad, 2018). More significantly, while such models are clearly preferable when mining low dimensional, structured data, some of the most medically consequential contributions of AI/ML systems have been based in the processing of complex, high-dimensional data by black box models (for instance in radiomics and medical imaging). Responding to this, Rudin has prospectively suggested that interpretation-aware methods such as in-building prototyping facilities can be integrated into even complex artificial neural nets (Chen et al., 2018; Li et al., 2018; Rudin, 2019), and, indeed, others have proposed that attention-based explainers be incorporated by design into model architectures of this sort (Choi et al., 2017; Park et al., 2016; Xu et al., 2018).[7]

The urgency of delivering rapid research responses to the COVID-19 pandemic puts a new kind of pressure on these emerging approaches to making complex, opaque models fit-for-purpose in supporting safety-critical decision-making. The development of new interpretability methods in clinical environments is likely to be put on the back burner, resulting in continued dependence on existing methodologies of explainability (for example, isolating the regions of interest in clinical imaging flagged by saliency maps or gradient class activation maps). As Wynants et al. demonstrate, however, even the essential process of properly annotating medical images during deep learning system design are not always well-executed in hasty research milieus (Wynants et al., 2020, p. 8), and time pressures placed on clinicians will challenge their capacities to thoroughly decipher and weigh up auxiliary explanation offerings. The outcomes of other potential applications of opaque model classes to unstructured, heterogenous data (or to combinations of this kind of data, say, free-text clinical notes, with EHRs) present explainability hurdles of their own. For instance, these may be explained by existing surrogate explanatory methods (like LIME or SHAP) that have been shown to have a spotty track record in generating accurate, reliable and faithful accounts of the determinant features driving black box predictions (Alvarez-Melis & Jaakkola, 2018; Leslie, 2019a; Mittelstadt, Russell, & Wachter, 2018; Molnar, 2019). We should note, additionally, that all of these options for supplementary, post hoc explanation support do not yet address more fundamental concerns that, even if partially explainable, some opaque models may still bury error-inducing faults or patterns of discrimination deep within their architectures that may manifest in unpredictable, unsafe, or inequitable processing outcomes.

---

[5] Another crucial component of building this sort of high performance, high interpretability model is the incorporation of domain knowledge to ensure that expert understanding of clinical conditions and underlying biological mechanisms is both informing feature selection and ultimately supporting the rationale behind the predictions—though a utilization of domain knowledge should not steer knowledge discovery away from exhaustive search of feature importance beyond existing insights (Gilvary, 2019; Jovanovic et al., 2016)

[6] Impactful contributions of the interpretability culture have included interpretable decision sets (Lakkaraju, 2016), generalized additive models (Lou et al., 2012), supersparse linear integer models (Rudin & Ustun 2016, 2018), certifiably optimal rule lists (Angelino et al., 2017), falling rule lists (Wang & Rudin, 2015), Boolean decision rules via column generation (Dash et al., 2018), and case-based reasoning (Bichindaritz & Marling, 2006; Bien & Tibrishani, 2011; Kim et al., 2016, adding criticism to prototypes).

[7] Many other strategies to design interpretable deep learning systems have also been investigated. For instance: (Bau et al., 2017; Wisdom et al., 2016; Zhang et al., 2018).

Although this enumeration of the difficulties faced by data scientists and AI/ML innovators is nothing new, a sense of urgency to confront them is. Taken together, the undercurrents of algorithmic bias, adverse data impact, and deficient process and outcome transparency are deep-rooted but open problems in data science that are presently made all-the-more challenging by the unprecedented pressures to tackle the pandemic. But these are problems with actionable solutions whose collective realization or evasion will be the historical axis that determines whether data science will be able to fulfil its massive potential to make a difference in the global fight against the virus. To set down a path towards this realization, data scientists will have to draw heavily upon the available moral-practical resources, existing knowledge, and sociotechnical self-understanding provided by current thinking in applied AI/ML ethics, social scientific approaches to data-driven technologies, and responsible research and innovation.

# 3 Digital contact tracing, solutionist lure or public health tool?

Before moving on to exploring the proper direction that responsible AI research and innovation should take, however, we would do well to investigate a controversial set of pandemic-related data-driven technologies. As is widely known, data-driven applications are being developed to speed up contact tracing and manage contagion through targeted health surveillance and individual tracking, as well as to enable personalised approaches to societal re-integration as social distancing measures are eased. These kinds of applications are triggering a complex set of ethical hazards that are only exacerbating the mounting challenges to autonomy, privacy, and public trust already faced globally by citizens caught in the crucible of a ubiquitously networked, big data society. Nevertheless, these kinds of human monitoring and tracking applications may prove crucial for managing the rapid asymptomatic and pre-symptomatic transmission of the disease and for mitigating some of its more punishing social and economic consequences (Ferreti et al., 2020).

## 3.1 The first wave of digital health surveillance in Asia

A first wave of such interventions, taking place in the Asian countries first struck by the virus, has largely been characterized by a combination of the macroscale exercise of social control and the centralized consolidation of personal and mobile phone tracking data. In China, technologists have built a non-compulsory but "use-to-move" AI application that integrates users' personal, health, travel, and location data with public health information about SARS-CoV-2 cases to produce individualized risk scores (stratified into "health code" levels of green, amber, and red). These determine who can access public spaces, shops, and public transport and who must be quarantined. The app, run through the prominent Alipay and WeChat platforms, is employed to monitor the movements of each of its roughly one billion users to ensure compliance and to keep continuous track of contacts (Calvo et al., 2020; Davidson, 2020; Mozur et al., 2020). Some reports out of China have been troubling. Not only may an AI-generated health code instantaneously turn from green to red without reason or explanation—as the algorithm behind the system is entirely opaque and made inaccessible to the public (Mozur et al., 2020)—a citizen caught traveling with a red code can be marked down in the country's social credit system to devastating personal and professional consequence (Zhang et al., 2020).

In Taiwan, a strategy of data integration similar to China's has been deployed that links the country's national health insurance database with its immigration and customs database to assist healthcare workers in identifying probable cases of COVID-19 infection (Wang et al., 2020). Taiwan has also used AI to monitor travellers, who are assigned risk scores based upon the origin and history of their travels and subsequently trailed through their mobile phones to confirm fulfilment of quarantine restrictions (Lee, 2020). Such a digital monitoring method can be heavy-handed; if a phone possessed by a quarantined user dies or is turned off, a visit from the police will soon follow (Lee, 2020). A different tack has been taken in South Korea, where the government is mining vast troves of CCTV, financial, and phone tracking data to reconstruct and publicize exhaustive—and potentially identifiable—logs of the movements and personal details of people who have tested positive for the virus (Zostrow, 2020). As has been noted by Nanni et al. (2020), such a method has marshalled data value to positive public health effect, while blatantly sacrificing patient privacy.

In contrast to South Korea, Singapore has taken a more consent-based and privacy-aware approach. It has implemented a Bluetooth-based proximity tracing system called TraceTogether—an opt-in decision-support application that helps public health officials to track down and communicate with the at-risk contacts of infected users (Bay et al., 2020). The TraceTogether app minimizes data collection by utilizing encrypted tokens that are exchanged between proximate users and then stored locally on their respective phones. Each user's non-personally identifiable tokens or "TempIDs" are issued by and stored on the server of the health authority, which also maintains a database of users' identifiable phone numbers. When individuals who have the app become ill with COVID-19, they are compelled by law to share their token exchange history with health officials, who are then able to use the central server to decrypt the tokens and compile a list of potentially infected users to contact (Bay et al., 2020; Cho et al., 2020).

## 3.2 The coming second wave and the prioritization of privacy

This Singaporean model has set the scene for the second, privacy-sensitive wave of data-driven surveillance and tracking technologies that has begun to form in Europe, the US and beyond. Here, the direction of innovation has largely been steered by concerns about intrusive data collection, use, and repurposing by centralized governmental or commercial infrastructures. Such anxieties have shaped debates around perceived trade-offs between priorities of privacy, individual liberties, and data protection, on the one hand, and, those of more collectively-oriented values such as protecting public health and community wellbeing, on the other. They have created an atmosphere of widespread apprehension which has led researchers and app developers to focus on finding technical solutions to the problem of optimizing privacy preservation while securing effective digital surveillance mechanisms.

Setting aside, for now, the ethical question of whether or not such a single-minded concentration on app-driven, technological solutions is justifiable given the plenitude of other relevant sociotechnical and practical factors at play (cf. O'Neill, 2020), two technical approaches have, so far, dominated the dash toward the development of privacy-preserving contact tracing technologies: GPS-based methods of co-localization tracing (Berke et al., 2020; Ferreti et al., 2020; Fitzsimons et al., 2020; Raskar et al., 2020 [though this presents hybrid features]; Reichert et al., 2020) and Bluetooth-based methods of proximity tracing (Bell et al., 2020; Brack et al., 2020; Canetti et al., 2020; Chan et al., 2020; Cho et al., 2020; CoEpi, 2020; COVID Watch, 2020; Hekmati et al., 2020; PEPP-PT, 2020; Tronosco, 2020; TCN, 2020). In the former, the GPS location histories of diagnosed carriers are de-identified and encrypted before they are shared with a backend server that allows for other users' apps to check whether or not they have crossed paths with the infected individual (Berke et al., 2020; Raskar, 2020). Proponents of this method argue that, despite some limitation of precision in determining collocation, its continuity with existing in-phone GPS tracking facilities will streamline the ease of its adoption, allowing for the magnitude of uptake necessary to lower COVID-19's $R_0$, its reproduction number, below 1 (some estimate 3/5 of a total population) and to consequently achieve herd immunity (Ferreti et al., 2020). Berke et al. maintain that the app's technology can be "integrated into partnering applications that already collect user location histories, such as Google Maps."

> These partner applications can then ask the user for the extra permissions and content for this system's use case. There are many such applications that already collect user location histories in the background. They often use this information to serve the user more relevant content and improve the user experience. However, this data collection more often serves private profit. Now, in the face of the COVID-19 pandemic, is the time for industry and researchers to come together

> and for the ubiquitous collection of location data to serve the public good (Berke et al., 2020, p. 11)

Among researcher and developers, however, there is increasing scepticism regarding the "significant privacy trade-offs" (COVID Watch, 2020) likely required in order for GPS-based methods of digital contact tracing to be functional. They have also flagged the correspondingly high computational burden placed on existing platforms by the cryptographic techniques needed to mitigate some of these issues (Bell et al., 2020; Chan et al., 2020). Another point of contention has been the accuracy limitations of location-centric methods—for instance, their inability to provide fine-grained recordings of interpersonal proximity and their lack of accurate functionality in certain buildings, subways, and multilevel dwellings. Such limitations call into question the capacity of GPS-based apps to measure human-to-human contacts with the degree of precision necessary to reflect medically defined specifications of disease exposure. These shortcomings have thus been taken to signal the advantages of Bluetooth-based techniques of proximity tracing (Canetti, 2020; Chan et al., 2020).

Unsurprisingly, Bluetooth-based tracing methods have now become the most likely data-driven technology to be applied to the health surveillance dimension of the current coronavirus pandemic in Europe and the US—a likelihood that has dramatically increased in light of an Apple/Google joint initiative to build a proximity tracing API for their three billion active mobile devices (Apple, 2020; Google, 2020). While human-in-the-loop proximity tracing technologies, like Singapore's TraceTogether, also use exchange and local storage of encrypted contact event tokens, many Anglo-European and American researchers are seeking to fully automate the Bluetooth-based contact tracing process so that all reliance on "trusted third parties" and data-consolidating central servers can be eliminated (Canetti et al., 2020; Chan et al., 2020; TCN, 2020; Tronosco, 2020).

In decentralized applications, users remain non-identifiable to each other from beginning to end of any contact tracing process. Data shared with central servers is minimized, and all contact detection, infection discovery, and risk computation is locally initiated and processed. When infection carriers are diagnosed, they receive health authority authorizations that then enable their anonymized contact histories to be uploaded onto a backend server. Meanwhile, the apps of other users periodically query this server to see if there are any contact matches. In the event that there are, each smartphone calculates a risk score and decides whether or not notification of its user is appropriate (esp. Tronosco, 2020). By taking humans out-of-the-loop through this type of comprehensively automated decentralization—a strategy whereby algorithmic models independently determine individual risk without the provision to users of any specific information backing or justifying the decision—the threats of adversary infrastructures that may violate privacy rights and infringe on data protections (whether they be governmental or commercial) are believed to be mitigated.

## 3.3 Public health priorities of past digital health monitoring interventions

Although the COVID-19 era emphasis on privacy and data protection has offered an important counterforce to the globally consequential threat of mass surveillance, questions remains as to whether this narrow focus on building airtight technological solutions has diverted attention away from some of the more salient underlying motivations and complexities that surround the introduction of digital contact tracing innovations during public health crises. Notably, the use of data-driven technologies to provide this kind of assistance was originally framed under the rubric of advancing digitally supported mobile health (mHealth) in the end of safeguarding community wellbeing (Danqua et al. 2019; Ha et al., 2016; Sacks et al., 2015; Reddy et al., 2015; Mendoza et al, 2014; World Bank, 2012; WHO, 2011). While several earlier approaches did prioritize privacy-preserving methods of digital contact tracing (Altuwaiyan, 2018; Prasad & Kotz, 2017; Shahabi 2015), the primary aim in pilot implementation studies and research interventions in this area was to optimize technological support for medical responses to ongoing epidemics. Thus, in Sacks et al. (2015), a smartphone-based mHealth tool was introduced to assist public health officials with Ebola surveillance and contact tracing in Guinea during the 2013-2015 epidemic. Though the tool faced significant adoption challenges, it "offered potential to improve data access and quality to support evidence-based decision-making for the Ebola response" (Sacks et al, 2015, p. 646).

Similar mHealth interventions were made by Ha et al. (2016) to assist with Tuberculosis contact tracing in Botswana in 2012 and by Danqua et al. (2019) during a 2015 Ebola outbreak in the Port Loko district of Sierra Leone. In the latter study, an Ebola contact tracing app was successfully deployed to streamline and support communication between contact tracers and the public health coordinators. Responding to the 2013 Dengue outbreak in Fiji, Reddy et al. (2015) introduced a GPS-based mHealth phone tracking tool that both helped public health officials to pinpoint infected areas and patients to become better informed about the symptoms of the disease and its treatments. The app also encouraged community-led support of public health efforts through cooperative involvement in reporting and identifying hotspots. Its creators concluded that the tool was "not going to replace physicians, however, it will greatly assist them in making their work easier in controlling disease outbreaks" (Reddy et al., 2015, p. 12).

### 3.4 The solutionist lures of automation all-the-way-down

A different, and potentially counterproductive, approach to digital contact tracing has developed in the context of the present coronavirus pandemic as normative emphasis has shifted from a stress on supporting medical response effectiveness to an emphasis on the extent to which the privacy-upholding expansion of automation can appease misgivings about state adversaries, mass surveillance, and function creep. The politics of public distrust may be purging the priority of the "public" from province of public health *per se* and providing an impetus to automation all-the-way down. While a data minimizing, privacy-preserving perspective on digital contact tracing is vital to its justifiability and societal acceptance, privacy-first apps that side-step the third-party, human-in-the-loop involvement of trusted contact tracers in investigation and risk determination should give us pause.

There are several reasons for this hesitation. First, a reliance on fully automated contact tracing methods for data collection and evaluation as well as for subsequent risk determination may betoken overconfidence in that data's accuracy, precision, and integrity and, consequently, in the reliability of the system that processes them. Common weaknesses in the integrity and quality of sensor data collected by digital devices (Ienca & Vayena, 2020) limit the likelihood that Bluetooth-based contact tracing technologies will be able to meet the high bar of functional requirements set by system designers themselves. In particular, barriers to measurement accuracy stemming from Bluetooth signals that fail to take account of glass windows, room dividers, product-lined shelves separating supermarket aisles, thin walls, mask-wearers, etc. (Fussell & Knight, 2020), raise questions as to whether this kind of contact tracing app can meet stated requirements such as "precision" (i.e. that "reported contact events must reflect actual physical proximity") and "integrity" (i.e. that "contact events are authentic") (Tronosco, 2020, p. 3).

To cut human contact tracers out of a public health process that is then bound to over-rely on fully automated tracing technologies is to preclude the application of common sense, context-awareness, and skilled judgement in remediating the data quality and integrity issues that will inevitably arise and in authenticating the veracity of data processing results. This difficulty is amplified in epidemiological settings, where the outcome-determinative gradients of encounters between infection carriers and at-risk individuals (close, casual, or transient) are often highly dependent on the contextual nuances of such factors as location and environment—nuances simply unavailable to the rigid algorithmic models behind contact tracing apps' risk calculations (Bay et al., 2020). For example, "short-duration encounters in enclosed spaces without fresh ventilation often constitute close contact, even if encounter proximity and duration do not meet algorithmic thresholds" (Bay et al., 2020, p. 6). Without the availability and use of common sense and human discernment, vital distinctions that might help public authorities avoid both false positives and false negatives will be lost. Such a judgment gap in the implementation of fully automated contact tracing systems suggests that the inferential brittleness of these apps may lead to ineffective or even deleterious “garbage-in-garbage-out” results. This would likely produce a generalized unease about adopting such technologies given the significant possibility that they will produce erroneous outcomes at the cost of either personal health or freedom of movement.

Furthermore, such a dislodging of human judgement raises the graver concern that taking humans out-of-the-loop may, in fact, contribute to the deterioration of social trust in the public health authorities charged with handling public health emergencies. Though crucial to take into consideration, the adversary assumption that eschews any trusted third party and motivates the comprehensive decentralization of digital contact tracing technologies is, perhaps, insufficiently attentive to the delicate role played by reciprocal relations of social trust and interpersonal responsibility in establishing and sustaining the fabric of shared confidence and mutual reliance that undergirds effective and community-involved public heath responses to public health crises.

The reason for this runs deep. Taken together, trust and responsibility have formed an implicit normative pillar of social order in the modern era. When individuals behave and act in ways that affect one another for better or worse, contemporary society binds them to the justifiability of their actions based upon reasonable expectations that, as rational agents, they will exercise good judgment in pursuing their objectives in ways that do not harm those around them and that are made accountable in virtue of such a ‘generalised expectancy’ (Rotter 1967). Securing such a nexus between the responsibility of each and the trust of others involves establishing a bedrock of situation-independent behavioural expectations between rational agents whereby mutually accountable performances can be universally assumed (Bauer & Freitag, 2018). Such a stable starting point for a free but orderly social coexistence has been variously called ‘basic trust’ (Erikson, 1959) and ‘generalised trust’ (Uslaner, 2002).

The problem with the complete automation of contact tracing is that it would do away with the architectonics of reasonable expectation that serve as an underpinning of generalised trust in the domain of public health. When crucial health decisions, such as a quarantine determination after an assessment of potential exposure, are taken out of the custody of responsible public health professionals, the kinds of reasonable expectations that anchor public trust (both in the institution and in the process) are likewise removed from the picture. Instead, a smartphone vibrates with an impersonal alert that perforce remains inexplicable to the potentially infected decision subject in its details and rationale. No reasonable expectations are involved inasmuch as one cannot, in effect, have these if there are no reasons on offer in the first place. And, when relations of reciprocal responsibility are consequently replaced by a *vox ex machina*, the unquestionable force of pre-emptory calculation leaves blind obedience to the algorithmic result as the only practicable option. Counterintuitively, this upshot of decentralization may have a kind of panoptical effect where a bloodless notification of infection risk on a mobile app punctuates a continuous dynamic of depersonalizing digital surveillance. At the negative extreme, this would mean that self-reinforcing mechanisms of social distrust end up optimizing privacy at the expense of creating conditions of deteriorated autonomy, social connectedness, and solidarity.

By contrast, in environments where research and innovation practices are organized around optimizing medical responses to public health emergencies and thus more directly oriented to the priority of societal wellbeing, digital contact tracing apps are seen as supporting evidence-based but compassion-driven human decision-making. For example, the creators of Singapore’s TraceTogether app have stressed the importance of a humane and human-centered approach: “Contact tracing involves an intensive sequence of difficult and anxiety-laden conversations, and it is the role of a contact tracer to explain how a close contact might have been exposed—while respecting patient privacy—and provide assurance and guidance on next steps” (Bay et al., 2020, p. 7). Here, the second-front design and deployment of a decision-supportive contact tracing technologies is understood to enable frontline contact tracers to “incorporate multiple sources of information, demonstrate sensitivity in their conversations with [citizens] who have had probable exposure to SARS-CoV-2, and help to minimise unnecessary anxiety and unproductive panic" (Ibid.).

## 3.5 Privacy, public health, and power

The contrast between the emerging privacy-first approach taken by proponents of fully automating digital contact tracing apps and the more public health-centered perspective instantiated in the Singaporean TraceTogether-supported method returns us to the debate around the seemingly unavoidable trade-offs between privacy and individual liberties, on one side, and community wellbeing and societal benefit, on the other. The difficulties faced at the extremes of the debate—at one end, the potential for radically centralized forms of health surveillance to lay waste to fundamental rights and freedoms, and at the other, the potential for radically decentralized forms to reinforce social distrust and to harm individual autonomy and interpersonal solidarity—should perhaps draw our attention to an additional factor that must also be considered. This is the issue of power as it relates to the use of data-driven technologies: What is the legitimate scope of the exercise of power during times of crisis and emergency? What are the real possibilities for its abuse or misuse, both by governments and by private companies, in regard to digital contact tracing and surveillance? What are the short- and long-term consequences of its impingement upon the digital organs that now sustain so much of our networked and connected private lives?

These questions highlight how the problem of power inexorably shades any consideration of the ethical challenges presented by digital contact tracing or beneficent health surveillance in contemporary big data society. Though the legitimacy of these sorts of data-driven technological interventions largely hinges on building reason-based public confidence in the appropriateness and justifiability of their employment, we do not, at present, live in a culture of public trust, when it comes to data collection, sharing, and use. The longstanding monetization of personal data by Big Tech companies has left members of society reasonably sceptical about how their data is being extracted and appropriated (Fourcade & Healy, 2013, 2017; Fuchs, 2010; Sadowski, 2019; Srnicek, 2016; Zuboff, 2015, 2019). After years of having algorithmically personalised services reach into their private lives to curate their tastes, nudge their behaviours, and steer their consumption, data subjects are sensibly on guard. Add to this the frightening but all-too-common instances, in many parts of the world, of intrusive governmental use of algorithmic targeting and manipulation for purposes of social control (Creemers, 2018; Roberts et al., 2019; Wright 2018, 2019), and it becomes easy to understand trepidation about the deployment of digital monitoring, tracking, and surveillance (Russell, 2019).

In the context of the second-front fight against COVID-19, attention to questions about power should key us in to the central importance of instituting regimes of responsible AI innovation in order to establish, and convince citizens about, the ethical justifiability, trustworthiness, and public benefit of such interventions. If data are to be legitimately marshalled through digital contact tracing, and health surveillance is to serve the purposes of community wellbeing, such innovations will have to be proportionate, socially licensed, and democratically governed. Normative AI regimes should ensure that research and innovation processes are reflective in anticipating ethical and societal impacts, that they are informed, from the start, by inclusive and collaborative deliberations on the balancing of potentially conflicting values, and that they are context-aware, domain-knowledgeable, and co-designed with the individuals and communities they affect. Such digital innovations will thus have to be explicitly values-driven, consent-based, and shaped by open public dialogue. Their processes of design and deployment will require transparency, continuous public oversight, rigorous pilot testing, reflective integration into wider public health strategies, and well-defined limitations. Developed responsibly, such technologies will have to be reasonably privacy preserving, compliant with human rights and responsible data management protocols, and subject to sunset and retirement provisions, which set clear and predefined constraints on their application to the present exceptional circumstances of the pandemic.

# 4 Five steps towards responsible AI innovation

A focus on responsible AI innovation, in the context of digital contact tracing and tracking apps, shows that it is essential not to fall prey to a tempting but false choice. This is between a sense that, in order to use these technologies, we must relinquish our fundamental rights and freedoms to the strengthening powers of the surveillance state and a sense that we must protect our privacy and individual liberties at the cost of pressing the full capacities of our data-driven technologies into the service of the public good. Both of these all-or-nothing alternatives fail to discern the potential of socially licensed innovation to function as a progressive counterforce to the excessive exercise of power. The potential rise of digital autocracies and AI-enabled totalitarian regimes, the abusive data grabs of state adversaries and profit-oriented commercial entities, the pre-emptive manipulation of human behaviour by platformed algorithmic infrastructures, these are real problems. But they are problems that modern free societies must combat by harnessing the democratic energies of open communication, public engagement, and collaborative value articulation. Drawing upon and strengthening the reflective, inclusive, and participatory character of practices of responsible innovation is, in fact, one of humanity's most effective instruments to accomplish this crucial task.

Even in the case of digital contact tracing and individual tracking, the cooperative steering and democratic governance of technology should, in this respect, be seen as a potential source of citizen empowerment and community-involving public health support rather than a fast track to despotic surveillance. In our time of pandemic, as leaders of nation-states take hold of extensive emergency powers, the deterioration of the rule of law, the possibility of the abuse of unchecked authority, and the potential for "surveillance creep" are hazards that merit sustained critical attention (Calvo, Deterding, & Ryan, 2020; French & Monahan, 2020). But, grim as they may be, these are political possibilities rather than societal inevitabilities, and they must be met head on by innovators, researchers, and citizens alike with the humane, communicative, and rational spirit of modern science. As nearly five centuries of the modern scientific method have shown, the open, dialogical, and consensus-based character of innovation processes are both a practical and epistemological necessity—a condition of possibility of the success of science itself (Appendix).

Considering all this, a starting point in practices of responsible innovation should be embraced as a first priority for those in the data science and AI/ML community, who are doing battle on the second front of the global struggle against COVID-19. Fortunately, the scientific community does not have to fly blind in figuring out how to meet these exigencies of ethical research and discovery. For almost half a century, concerted efforts to flesh out responsible ways of pursuing the design and use of increasingly powerful technological tools have been made in areas ranging from bioethics (Beauchamp & Childress 2001; DHEW, 1974; Kuhse & Singer, 2009) and responsible research and innovation (RRI) (Hellström, T., 2003; Owen, Macnaghten, & Stilgoe, 2012; Von Schomberg, 2011, 2013, 2019); to applied ethics (May & Delston, 2016; Singer, 1980, 1986), science and technology studies (STS) (Jasanoff, 2012, 2016; Nissenbaum, 2001; Sengers et al., 2005; Star, 1999), and, more recently, digital and AI/ML ethics (EPSRC, 2011; Floridi, 2010; Floridi & Cowls, 2019; Jobin, Ienca & Vayena, 2019; Leslie, 2019a; Zeng et al., 2019).

We might do well, then, to turn to this body of research as a way to start upon a much longer journey toward creating a culture of responsible innovation in the data science and AI community. For this reason, I want to move now to proposing five steps that should be taken in order to responsibly bring the insights of data science and the tools of AI/ML to bear on the wide range of biomedical, epidemiological, and socio-economic problems raised by the coronavirus pandemic. When incorporated into research and innovation processes from the start, these best practices will not only enhance the quality of research and discovery without adding undue burdens, they will improve the quality of outcomes and results. To put it simply, *responsible* data science is *good* data science—data science *with* and *for* society and worthy of public trust.

## *Step I: Open science and share data responsibly*

**Open science** and **open research** build public trust through reproducibility, replicability, transparency, and research integrity (European Commission, 2014; McNutt, 2014; NAS, 2018, 2019; Nosek et al., 2015; The Turing Way, 2020). The cooperative and barrierless pursuit of scientific discovery accelerates innovation, streamlines knowledge creation, fosters discovery through unbounded communication, and increases the rigour of results through inclusive assessment and peer review (Fecher & Friesike, 2014). Opening models and research procedures to expert assessment, oversight, and critique allows for rapid error and gap identification and catalyses the improvement of results. Moreover, reproducible and replicable research that is made accessible to all helps create confidence across society in the validity of scientific work.

The global reach of open research to an unbounded scientific community is especially important in the battle against COVID-19. The coronavirus pandemic is a species-level crisis, and so the scope and cooperative reach of the practices of scientific ingenuity that seek to redress it should also be global and inclusive. Managing the spread of the infection effectively will involve bolstering the knowledge as well as the control and mitigation strategies of every nation great and small.

A crucial constituent of this global effort is **responsible data sharing.** While the first critical step in this direction is to *open up data* so that research can be reproduced and re-used, datasets can be iteratively improved, and investments of time and research funding can feedforward to keep benefitting the public good (Borgman, 2015; Burgelman et al., 2019; Molloy, 2011; Piwowar et al., 2011; Tenopir et al., 2011; Whitlock, 2011), the concept of "open data" itself must be bounded and qualified (Dove, 2015; Jasanoff, 2006; Leonelli, 2019). Data sharing does not occur in a sociocultural, economic, or political vacuum but is rather situated amidst an interconnected web of complex social practices, interests, norms, and obligations. This means that those who share data ought to practice critical awareness of the moral claims and rights of the individuals and communities whence the data came, of the real-world impacts of data sharing on those individuals and communities, and of the practical and sociotechnical barriers and enablers of equitable and inclusive research.

First and foremost, data sharers have a responsibility to serve the interests of wider society through the ethically piloted advancement of science, while simultaneously protecting the privacy and interests of affected data subjects. Accessible, high-quality, and well-archived data are the most critical ingredients in the progress of data scientific insights and AI/ML technologies, but responsibly opening data also involves privacy optimised, impact aware, and security compliant data sharing. These two components can be seen as complimentary: Properly managed accessibility and maximal data integrity allow for trusted data to be more freely circulated among an ever-widening circle of responsible researchers so that results can be replicated, and new, societally beneficial insights produced.[8] Responsible research that moves in this direction should refer to well-established protocols for responsible data management like those of the FAIR data principles (findable, accessible, interoperable and reusable data) (Wilkinson et al., 2016), trusted digital repositories (ISO 16363), Criteria for Trustworthy Digital Archives (DIN 31644), and the Data Archiving and Networked Services' CoreTrustSeal.

Data scientists and AI researchers, who are tackling COVID-19 should also take heed of the higher demands for **data integrity** in safety-critical and highly regulated environments like healthcare. Data integrity, in this vein, can be understood as those dimensions of responsible data governance that safeguard trustworthiness across the entire data lifecycle from collection and correction through processing and retention. A useful framework for responsible end-to-end data governance is the "five safes" published by the UK's Office for National Statistics (Desai, Ritchie, & Welpton, 2016). The "five safes" aim to ensure that data is used for a morally and legally justifiable purpose and for the public benefit (safe projects), that researchers are well-trained and can be trusted to use the data appropriately (safe people), that the data is reliably de-identified (safe data), that access to the data is managed in a secure and situation appropriate way (safe settings), and that research outputs are non-disclosive and do not provide opportunities for re-identification (safe outputs). Additionally, a high bar for standards of data integrity can be found in the "ALCOA plus" principles, which have been condoned and described in helpful guidance on data integrity (in the context of pharmaceuticals and medical devices) produced by the World Health Organisation and by the UK's Medicines and Healthcare products Regulatory Agency (WHO, 2014; MHRA, 2018).

[8] A notable initiative in this direction has already been made by the Research Data Alliance (Berman & Crosas, 2020). In the area of health data, the UK's Health Data Research UK (HDR UK) is also making major strides forward in institutionalizing responsible data sharing as is the Coleridge Initiative.

The special responsibilities shouldered by researchers who are trying to apply responsible data sharing practices in a global public health crisis has been broached in the World Health Organisation's 2015 consultation, *Developing global norms for sharing data and results during public health emergencies* (Modjarrad, 2016). Here, the WHO stresses that "timely and transparent pre-publication sharing of data and results during public health emergencies must become the global norm" (WHO, 2015, intro.). Moreover, it affirms that opting in to the sharing of data and data analyses must be treated as a default practice and a moral obligation:

> Every researcher that engages in generation of information related to a public health emergency or acute public health event with the potential to progress to an emergency has the fundamental moral obligation to share preliminary results once they are adequately quality controlled for release. The onus is on the researcher, and the funder supporting the work, to disseminate information through pre-publication mechanisms, unless publication can occur immediately using post-publication peer review processes (WHO, 2015, para. 2).

Notwithstanding the WHO's endorsement of open research and responsible data sharing, authors of the background briefing prepared for the 2015 consultation identified several barriers to information sharing (Goldacre et al., 2015) that also figure in the context of the COVID-19 pandemic. These include issues related to information governance and data protection when ambiguities arise regarding informed consent and the confidentiality of potentially re-identifiable personal data, tensions between the need to share results rapidly and risks of inaccurate information doing harm in clinical environments, legacies of proprietary protectionism and the chilling effects of motivations to hoard data in the ends of academic publication priority, and delays in data sharing caused by the lengthy peer review processes involved in scientific journal publication.

Though many of these issues may be addressed through deliberate attitude change and the institution of governance regimes that ensure transparency and accountability, other barriers to responsible data sharing are rooted in more intractable social formations such as underlying global inequalities and territorially and regionally variant political priorities that undermine federated, international approaches to addressing public health emergencies through open research. These are presenting scientists and innovators combating SARS-CoV-2 on the global plane with difficulties that are less immediately soluble but that should nevertheless be kept in view.

To take the issue of political priorities first, fears of outbreak-related reputational damage, migration and trade restrictions, widespread social stigma, damage to financial markets, and exposure of national security vulnerabilities may lead countries, political leaders, and state-controlled institutions to dissemble data and to clamp down on information dispersion. Varying instances of this occurred in the 2003 SARS CoV outbreak, in the 2009 H1N1 influenza pandemic, and in recent cholera outbreaks (Briand, Mounts, & Chamberland, 2011; Goldacre et al., 2015; Huang, 2003). Despite the explicit reporting and information-sharing provisions in the WHO's 2005 *International Health Regulations*, the high economic and geopolitical stakes of global public health emergencies can motivate political actors to engage in obstructive behaviours that prevent forthright and unhindered data dissemination.

Heeding these possibilities, data scientists and AI innovators must prioritize boots-on-the-ground communication with the researchers, clinicians, and domain experts, who are directly involved in responding to and gathering data about the COVID-19 pandemic. More than that, innovators should bear in mind these political factors when they critically assess changes in the data landscapes as the current global public health crisis runs its course.

A second barrier to responsible data sharing, to which data scientist and AI/ML innovators should pay close attention, originates in long-standing dynamics of global inequality that may undermine reciprocal sharing between research collaborators from high-income countries (HICs) and those from low/middle income countries (LMICs). Given asymmetries in resources, infrastructure, and research capabilities, data sharing between LMICs and HICs, and the transnational opening of data, can lead to inequity and exploitation (Bezuidenhout et al., 2017; Leonelli, 2013; Shrum, 2005). For example, data originators from LMICs may put immense amounts of effort and time into developing useful datasets (and openly share them) only to have their countries excluded from the benefits of the costly treatments and vaccines produced by the researchers from HICs who have capitalized on such data (Goldacre et al., 2015).

Moreover, data originators from LMICs may generate valuable datasets that they are then unable to independently and expeditiously utilize for needed research, because they lack the aptitudes possessed by scientists from HICs who are the beneficiaries of arbitrary asymmetries in education, training, and research capacitation (Bull et al., 2015; Merson et al., 2015). This creates a two-fold architecture of inequity wherein the benefits of data production and sharing do not accrue to originating researchers and research subjects, and the scientists from LMICs are put in a position of relative disadvantage vis-à-vis those from HICs whose research efficacy and ability to more rapidly convert data into insights function, in fact, to undermine the efforts of their disadvantaged research partners (Bezuidenhout et al., 2017; Crane, 2011).

This challenge of misshapen reciprocity brings out a deeper issue pertaining to the framing of the desideratum of open data. While the challenge of overcoming the problem of global digital inequality in the era of data-driven innovation has often been approached under the rubric of traversing the "digital divide" through more equitable provision of the resources needed to access information and communication technologies (ICTs), such a perspective neglects the enabling conditions of the globally diverse and disparately resourced *practices of innovation* that are needed to convert such technological resources into insights and applications. It is important, that is, to quarry beneath the issues of resource availability and allocation of ICTs, which have largely framed the impetus to opening data, and to concentrate as well on what Bezuidenhout et al. refer to as "the infrastructural, social, institutional, cultural, material and educational elements necessary to ensure the realization of openness" (Bezuidenhout et al., 2017, p. 465).

On this view, in redressing the barriers of inequality that hamper the responsible opening of data, emphasis must be placed on "the social and material conditions under which data can be made useable, and the multiplicity of conversion factors required for researchers to engage with data" (473). Equalizing know-how and capability is a requisite counterpart to equalising access to resources, and both together are necessary preconditions of ethical data sharing. With this in mind, data scientists and AI/ML innovators engaging in international research collaborations should focus on forming substantively reciprocal partnerships where capacity-building and asymmetry-aware practices of cooperative innovation enable participatory parity and thus greater research equity.

### *Step II: CARE & Act through Responsible Research and Innovation (RRI)*

This demand for researchers to be responsive to the material and social preconditions of responsible innovation practices, reminds us of the wider practical purview of RRI. An RRI perspective provides researchers and innovators with a vital awareness that all processes of scientific discovery and problem-solving possess sociotechnical aspects and ethical stakes. Rather than conceiving of research and innovation as independent from human values, RRI regards these activities as morally-implicated social practices that are duly charged with a responsibility for critical self-reflection about the role that such values play in discovery, engineering, and design processes and in considerations of the real-world effects of the insights and technologies that these processes yield.

The RRI view of 'science with and for society' has been transformed into helpful general guidance in such interventions as EPSRC's 2013 AREA framework and the 2014 Rome Declaration. These emphasize the importance of anticipating the societal risks and benefits of research and innovation through open and inclusive dialogue, of engaging with affected stakeholders as a means to co-creation at all stages of the design, development, and deployment of emerging technologies, and of ensuring transparent and accessible innovation processes, products, and outcomes (Owen, 2014; Owen, Macnaghten, & Stilgoe, 2012).[9] The AREA Framework (Anticipate, Reflect, Engage, Act) is a handy tool to continuously sense check the social and ethical implications of innovation practices. Adding to this the priority of contextual considerations, we have the CARE & Act Framework:

> **Consider context**—think about the conditions and circumstances surrounding research and innovation. Focus on the practices, norms, and interests behind it. Take into account the specific domain in which it is situated and reflect on the concrete problems, attitudes, and expectations that derive from that domain;
>
> **Anticipate impacts** – describe and analyze the impacts, intended or not, that might arise. This does not seek to predict but rather to support an exploration of possible risks and implications that may otherwise remain uncovered and little discussed;
>
> **Reflect on purposes** – reflect on the goals of, motivations for, and potential implications of the research, and the associated uncertainties, areas of ignorance, assumptions, framings, questions, dilemmas and social transformations these may bring;
>
> **Engage inclusively** – open up such visions, impacts and questioning to broader deliberation, dialogue, engagement and debate in an inclusive way. Embrace peer review at all levels and welcome different views; and
>
> **Act responsibly** – use these processes to influence the direction and trajectory of the research and innovation process itself. Produce research that is both scientifically and ethically justifiable. (EPSRC, 2013, amended and expanded)

The CARE & Act Framework provides an actionable way to integrate anticipatory reflection and deliberation into research and innovation processes, while also emphasizing that an earlier stage-setting step must be taken to enable such an approach. Building this bridge from context to anticipation, reflection, and engagement is crucial. A solid understanding of innovation context is a precondition of effective anticipatory reflection inasmuch as it provides access to the key domain- and use-case-specific needs, desiderata, obligations, and expectations that frame pre-emptory considerations of the potential risks and impacts of any given research and innovation project. For instance, domain-situated knowledge of an AI system's operating environment will yield useful information about relevant industry standards and norms, organizational and public expectations, and outcome-influencing social factors and circumstantial exigencies. By taking contextual aspects like these into account, researchers and innovators will be better able to weigh up risks and impacts, to elicit the design and implementation requirements that address or mitigate them, and to take deliberate design-time actions to meet these requirements.[10]

[9] It would be helpful to note that there has been a high degree of critical self-reflection in RRI about limitations in the generalizability and succinctness of its framing of values-based research and discovery. For instance, issues have been raised about its naïve grouping of the classes of research and innovation, about inexorable definition disagreements regarding action-orienting values, and about the potential for abuse and misuse of public engagement methods (Jirotka et al., 2017).

There is one other component of RRI's capacity to build the bridge from context to anticipation, reflection, and engagement that is important to mention. Fruitful efforts to integrate the embodied, interactive, and pragmatic perspective of human-computer interaction (HCI) scholarship into RRI have helped to highlight the importance of contextual self-awareness and situational responsiveness in responsible innovation practices (Eden et al., 2013; Grimpe et al., 2014; Jirotka et al., 2017; Stahl & Coeckelbergh, 2016). In particular, reflexivity and anticipation are seen, from this standpoint, as concretely enacted amidst the needs, opportunities, and problems of the particular communities of practice in which innovators and researchers are embedded (Eden et al., 2013). This means that contexts of innovation are animated for these innovators and researchers through their responsiveness to real-world challenges and to the continual demands of collaborative problem-solving. Such a de-idealized mode of "reflection-in-action" (Eden et al., 2013, p. 2972) consequently enables practices of RRI to stay warm-blooded and agile as scientists and innovators face the novel ethical difficulties posed by unforeseen problems and unknown unknowns.

To tackle COVID-19 responsibly, data science researchers and AI/ML innovators will have to marshal this agility and situational responsiveness as they cope with the innovation context of the present global health crisis. Helpful resources for gaining a general working understanding of this contextual dimension can be found in the World Health Organisation's *Guidance for Managing Ethical Issues in Infectious Disease Outbreaks* (2016), in the Council for International Organization for Medical Sciences' *International Ethical Guidelines for Health-Related Research Involving Humans* (2016), and in the Nuffield Council on Bioethics' *Research in global health emergencies: ethical issues* (2020). Against the specific backdrop of data science and AI innovation, the following non-exhaustive list of contextual considerations may help orient anticipatory reflection within the frame of the coronavirus pandemic:

> **Magnified harmful effects on vulnerable and disadvantaged communities**- As we are already seeing in the devastating impact of the SARS-CoV-2 outbreak on communities of color and impoverished social groups, the pandemic is disproportionately affecting members of our society who are subject to structural legacies of disadvantage that put them at greater risk than others. When designing and developing innovation, researchers must take heed of these circumstances of vulnerability and disadvantage (MacIntyre & Travaglia, 2015). They must focus on protecting those who are most at risk and on ensuring that technological interventions purposefully yield societally equitable outcomes.
>
> **Disruption of public order and social, moral, political, and legal norms**- The governmental exercise of emergency powers and the urgency of producing swift and effectively scaled responses to public health crises can disrupt public order, societal norms, and the rule of law. This may occasion abrupt and wide-scale social changes, which subsequently have deleterious or regressive long-term consequences. For instance, if pursued without predefined limitations, the enforcement of censorship and surveillance measures to protect the public during an outbreak could shift norms of public acceptability and legal protections away from the safeguarding of civil liberties and fundamental rights and freedoms. Scientists and innovators should proceed with vigilance in analyzing the protracted effects of the innovations they produce.

[10] This context-aware and anticipatory approach has been developed in the area of argument-based assurance of safety-critical digital technologies. "Assurance cases" or "safety cases" provide an integrative and process-based platform for ensuring that the properties needed to fulfil the high-level normative goals of a computational system and to mitigate its anticipated risks are translated into design actions and documented as interrelated claims, arguments, and evidence. Consolidated standards for system and software assurance include the ISO/IEC/IEEE 15026 series and the Object Management Group's Structured Assurance Case Metamodel (SACM), and various assurance platforms exist such as Goal Structuring Notation (GSN), the Claims, Arguments and Evidence Notation (CAE), and Dynamic Safety Cases (DSC). For further background, see: Ashmore et al., 2019, Bloomfield & Bishop, 2010; Bloomfield & Netkachova, 2014; Denny et al., 2017; Ge et al., 2012; Health Foundation, 2012; Kelly, 1998, 2003; Kelly & Weaver, 2004; and Picardi et al., 2019.

**Compromised consent and decision-making-** Public health crises put affected individuals and communities as well as frontline care providers under conditions of extreme duress, urgency, and distress (BMA, 2020; Nuffield, 2020; WHO, 2016). Those who are stricken with infection, or have sick family members, have to cope with uncertainty, suffering, fear, and powerlessness—all of which can compromise the processes of assessment, deliberation, and judgement that are required for the provision of informed consent. Likewise, in overwhelmed clinical environments, healthcare professionals are faced with constant demands to render critical decisions under conditions of incomplete information, extreme urgency, uncertainty, and disorder. Scientists and innovators must carefully take into account the distressed circumstances of those impacted by their research, and they must, where possible, prioritize ways of gaining informed and voluntary consent that accommodate these challenges, while also respecting the dignity of every person, recognizing their unique hardships, and taking into account the reasonable expectations of impacted individuals (consistent with Barocas & Nissenbaum, 2014; Nissenbaum, 2009). Likewise, innovators who are designing AI/ML decision-support systems for distressed clinical environments must take into consideration their distinctive implementation needs.

## *Step III: Adopt ethical principles to create a shared vocabulary for balancing and prioritising conflicting values*

In our pluralistic and culturally diverse world, resolving ethical dilemmas is often dependent on building inclusive and well-informed consensus rather than appealing to higher authorities or to the say-so of tradition. This need for consensus-building is especially crucial in the context of AI/ML innovation, where circumstances often arise in which ethical values come into tension with each other. For instance, there may be situations (such as with digital contact tracing) in which the use of data-driven technologies may advance the public interest only at the cost of safeguarding certain dimensions of privacy and autonomy. Trade-offs, in cases like these, may be inevitable, but, regardless, the choices made between differing values should occur through a medium of equitable deliberation, mutual understanding, and inclusive and knowledgeable communication.

To this end, it is especially important to set up procedural mechanisms that enable reciprocally respectful, sincere, and open dialogue about ethical challenges. These mechanisms should help conversation participants speak a common language so that, when an innovation project's potential social and ethical impacts are being assessed and re-assessed, diverging positions can be weighed and reasons from all affected voices can be heard, understood, and suitably considered. This can be accomplished by adopting common ethical principles from the outset to create a shared vocabulary for informed dialogue about balancing conflicting values.

There is, however, an obvious and important stumbling block that must be dealt with by this point of view. Amidst the kaleidoscopic plurality of modern social life, no fixed or universally accepted list of ethical values could pre-reflectively provide such a common starting point. Researchers in AI/ML ethics have therefore had to take a more pragmatic and empirically-driven position, in proposing basic values, that begins by considering the set of real-world problems posed by the use of the AI/ML and data-driven technologies themselves. These hazards include the potential loss of human agency and social connection in the wake of expanding automation, harmful outcomes that may result from the use of poor quality data or poorly designed systems, and the possibility that entrenched societal dynamics of bias and discrimination will be perpetuated or even augmented by data-driven AI/ML technologies which tend to reinforce existing social and historical patterns.

In responding to such hazards, dozens of frameworks in AI/ML ethics have, over the past few years, more or less coalesced around a set of principles originating in both bioethics and human rights regimes (for example, Floridi & Clement-Jones, 2019; HILEG, 2019; IEEE, 2018; OECD, 2019; Université de Montréal, 2017; ).[11] The UK Government's official public sector guide to safe and ethical AI has consolidated these into four "SUM values"—values that aim to *support*, *underwrite*, and *motivate* a responsible and reflective AI/ML innovation ecosystem and that are anchored in ethical concerns about human empowerment, interactive solidarity, individual and community wellbeing, and social justice (Leslie, 2019a). These are:

**Respect** the dignity of individuals as persons:

- Ensure the abilities of individuals to make free and well-informed decisions about their own lives
- Safeguard their autonomy, their power to express themselves, and their right to be heard
- Value the uniqueness of their aspirations, cultures, contexts, and forms of life
- Secure their ability to lead a private life in which they are able to intentionally manage the transformative effects of the technologies that may influence and shape their development
- Support their abilities to fully develop themselves and to pursue their passions and talents according to their own freely determined life plans

**Connect** with each other sincerely, openly, and inclusively:

- Safeguard the integrity of interpersonal dialogue and connection
- Protect human interaction as a key for trust and empathy
- Use technology to foster this capacity to connect so as to reinforce reciprocal responsibility and mutual understanding

**Care** for the wellbeing of each and all:

- Design and deploy AI to foster and to cultivate the welfare of all stakeholders whose interests are affected by their use
- Do no harm with these technologies and minimise the risks of their misuse or abuse
- Prioritize the safety and the mental and physical integrity of people when scanning horizons of technological possibility, conceiving of, and deploying AI applications

**Protect** the priorities of justice, social values, and the public interest:

- Treat all individuals equally and protect social equity
- Use digital technologies to support the protection of fair and equal treatment under the law

[11] Crucially, these ethical principles have arisen in both bioethics and human rights regimes as moral claims that have responded directly to tangible, technologically-inflicted harms and atrocities. In a significant sense, that is, both traditions emerged out of concerted public acts of resistance against violence done to disempowered or vulnerable people. Whereas human rights has its origins in efforts to redress the well-known barbarisms and genocides of the mid-twentieth century, in the case of bioethics, its emergence tracked the public exposure in the 1960's and 1970's of several atrocities of human experimentation (such as the infamous Tuskegee syphilis experiment), where it was discovered that members of vulnerable or marginalised social groups had been subjected to the injurious effects of institutionally run biomedical experiments without having knowledge of or giving consent to their participation (Kuhse & Singer, 2009; Leslie et al., 2020). While a longer discussion of this is out of the scope of this paper, it is notable that the provisional universalism of AI/ML ethics principles is rooted in a kind of moral grammar that underlies acts of resistance against those who have perpetrated social injury (Honneth, 2007).

- Prioritise social welfare, public interest, and the consideration of the social and ethical impacts of innovation in determining the legitimacy and desirability of AI technologies
- Use AI to empower and to advance the interests and well-being of as many individuals as possible
- Think big-picture about the wider impacts of the AI technologies you are conceiving and developing. Think about the ramifications of their effects and externalities for others around the globe, for future generations, and for the biosphere as a whole

These SUM values form the basis of the Stakeholder Analysis component of the NHSx's *Code of Conduct for Data-Driven Health and Care Technology*. They are intended as a launching point for open and inclusive conversations about the individual and societal impacts of AI/ML innovation projects rather than to provide a comprehensive inventory of moral concerns and solutions. At the very outset of any AI/ML project, these should provide the normative point of departure for collaborative and anticipatory reflection, while, at the same time, allowing for the respectful and interculturally-sensitive inclusion of other points of view.

It should be noted, here, that circumstances of a public health crisis such as the COVID-19 pandemic may place processes of deliberatively balancing and prioritizing conflicting or competing values under extreme pressure to yield decisions that generate difficult trade-offs between equally inviolable principles. In all cases, though there may be no *a priori* prescription or moral formula to determine such decisions in advance, research and innovation projects in data science must remain lawful and bound by obligations codified in existing international human rights agreements (WHO, 2016) and data protection law (the GDPR, paradigmatically). In this respect, the *Siracusa Principles on the Limitation and Derogation Provisions in the International Covenant on Civil and Political Rights* (Siracusa, 1985), provide a helpful reference point for considerations of the placement of permissible limitations on fundamental rights and freedoms in emergency situations where certain trade-offs are unavoidable to achieve legitimate objective interests. These affirm that any such restrictions should be a last resort after all other possible alternatives (which would have achieved the same outcome less intrusively) are exhausted, and that such restrictions should be legal, proportionate, reasonable, reviewable, evidence-based, and equitably executed (Boggio et al., 2008; Todrys, Howe & Amon, 2013).

### *Step IV: Generate and cultivate public trust through transparency, accountability, and consent*

The ultimate success of any AI/ML innovation project undertaken to combat COVID-19 will not only hang on the quality and performance of the product. It will also rest on whether or not a degree of public confidence in the safety and responsibility of the innovation has been established that is sufficient to foster its adoption by the healthcare community and society at large. Three key preconditions of trustworthy innovation deserve special attention.

First, all AI/ML innovation projects should proceed with **end-to-end transparency** to establish both that design, discovery, and implementation *processes* have been undertaken responsibly and that *outcomes* are appropriately explainable and can be conveyed in plain language to all affected parties. Research undertaken to combat the SARS-CoV-2 outbreak should *ceteris paribus* occur as openly as possible. It should be carried out in a way that demonstrates to the public that innovation processes and products are ethically permissible as well as fair, safe, and worthy of trust (ACM, 2017; Leslie 2019a). This entails the adoption of best practices mechanisms for responsible data sharing and for the assurance of data integrity such as the FAIR data and ALCOA plus principles mentioned above. Moreover, as our discussion of Wynants et al. (2020) suggested, research practices and methodological conduct should be carried out deliberately, transparently, and in accordance with recording protocols that enable the reproducibility and replicability of results. For prediction models, the documentation protocols presented in Transparent Reporting of a Multivariable Prediction Model for Individual Prognosis or Diagnosis (TRIPOD) is a good starting point for best conduct guidelines in reporting (Collins et al., 2015; Moons et al., 2015).

Researchers and innovators should likewise ensure that the results of their AI/ML models are reasonably and appropriately intelligible to users and affected individuals. Interpretable models and results will be a crucial factor in the adoption of AI/ML decision-support systems in clinical environments. They will also enable more effective and evidence-based assurance that AI/ML systems will operate safely, reliably, robustly, and equitably. Though this still remains something of a difficult issue for the complex, opaque classes of AI/ML algorithms, researchers and innovators should nevertheless prioritize the interpretability and explainability of their models from the start of their projects and, where applicable, maximize the accuracy and fidelity of any supplementary explanation methods they use to access the rationale of the complex models they deploy. They should also prioritize the use of interpretable methods, when structured data with meaningful representations are being utilized and pursue diligent techniques of iterative knowledge discovery as well as sufficient consultation with domain experts (Gilvary, 2019; Rudin, 2019).

A helpful reference point for this component of outcome transparency can be found in *Explaining decisions made with AI*, a guidance recently published by the UK's Information Commissioner's Office and The Alan Turing Institute. This guidance takes a holistic, end-to-end, and context-based approach to building AI/ML systems that are explainable-by-design. It focuses on the importance of tailoring both design-time and run-time strategies of producing understandable results to each model's specific use-case and practical context.
This vital contextual aspect includes the specific subfield or area in which the clinical end-user operates, and the individual circumstances of the person receiving the decision. The guidance stresses a values-based approach to the governance of AI/ML explanations, presenting four principles of explainability that steer the recommendations it proposes: be transparent, be accountable, consider context, and reflect on impacts.

Building off these, it identifies a range of different explanation types, which cover various facets of an explanation, such as explanations of who is responsible, explanations of the rationale that led to a particular decision, explanations of how data has been collected, curated, and used, and explanations of measures taken across an AI/ML model's design and deployment to ensure fair and safe outcomes (Leslie & Cowls, 2020). Finally, it emphasizes that, in every individually impacting case, the statistical generalizations that underlie the rationale of any decision-support system's output should be translated into plain, socially meaningful language and applied by the end-user or implementer with due regard for the concrete life circumstance of the affected decision subject.

A second precondition of trustworthy innovation is accountability. All AI/ML innovation projects should proceed with **end-to-end accountability** to ensure both that humans are answerable for the parts they play across the entire AI/ML design, discovery, and implementation workflow and that the results of this work are traceable from start to finish. Diligent accountability protocols that are put in force across the AI/ML lifecycle will ensure public confidence that innovation processes prioritise patient and consumer interests from beginning to end. Members of civil society, domain experts, and other relevant stakeholders should also be included in the AI/ML workflow through the institution of independent advisory consortia, which function as sounding boards as well as sense-checks and oversight mechanisms throughout innovation processes.

Finally, these regimes of transparency and accountability should facilitate **informed community and individual consent** that reflects the contexts and reasonable expectations of affected stakeholders**.** Trust-building through community consultation should be utilized to foster the development of equal and respectful relationships—true partnerships—among researchers, healthcare professionals, and affected individuals and communities (Wright et al., 2020). Furthermore, public buy-in should come both from the groups in wider society that are impacted by the products of AI/ML innovation projects and from each individual who is directly affected by the use of these products. This can be achieved, on this broader scale, through effective *ex ante* public communication of the scope and nature of the AI/ML innovations undertaken. Such public engagement should provide non-technical synopses of the research as well as summaries of the measures taken across the project lifecycle to ensure safe, ethical, equitable, and appropriately explainable outcomes.

### *Step V: Foster equitable innovation and protect the interests of the vulnerable*

Even before the COVID-19 pandemic, vulnerable and historically disadvantaged social groups were especially in peril of being harmed by or excluded from the benefits of data-driven technologies (Barocas & Selbst, 2016; Eubanks, 2018; Gianfranceso et al., 2018; Noble, 2018). Patterns of social inequity, marginalisation, and injustice are often "baked in" to the data distributions on which AI/ML systems learn. Over the past decade, a growing body of fairness-aware and bias-mitigating approaches to AI/ML design and use has been bringing many of these issues out into the open both in terms of academic research (for helpful surveys: Friedler et al., 2019; Grgic-Hlaca et al., 2018; Mehrabi et al., 2019; Romei & Ruggieri, 2013; Verma & Rubin, 2018; Žliobaitė, 2017) and in terms of practically applicable user interfaces (several tools for fairness-aware design and bias auditing have been created such as University of Chicago's Aequitas open source bias audit toolkit for machine learning developers, TU Berlin's Datasets and software for detecting algorithmic discrimination, and IBM's Fairness 360 open source toolkit).[12] This increasing focus on issues of bias and discrimination has brought needed attention to the deep-rooted dynamics of dataset discrimination that are in peril of perpetuating many existing health inequities. Such dynamics have been evidenced, for example, in studies that have shown patterns of misclassified risk assessment of inherited cardiac conditions in black Americans for reason of their lack of representation in genetic datasets (Manrai et al., 2016), information disparities across racial, ethnic, and ancestral subgroups about clinically relevant genetic variants in the Genome Aggregation Database (Popejoy et al., 2018), biased clinical risk assessment of atherosclerotic disease due to the overrepresentation of white patients in the Framingham Risk Factors cardiac evaluation tool (Gijsberts et al., 2015), and information gaps in the capacity of decision support tools to pick up diagnostic and treatment relevant signals in EHRs from vulnerable patient subgroups, who have irregular or limited access to healthcare (Arpey, 2017; Gianfancesco, 2018; Ng et al., 2017).

Though these instances highlight the importance of scrutinizing rapidly proliferating COVID-19 datasets for representativeness, balance, and inclusion of relevant information about all affected social groups across the demographic whole, the prevalence of health inequities they indicate call attention to other potential sources of pandemic-related digital discrimination. All-too-often, vulnerable or socioeconomically disadvantaged stakeholders are subject to material conditions, which make access to potentially beneficial digital technologies unavailable (Cahan et al., 2019; Weiss et al., 2018). Those who design digital apps used for contact tracing (and all other proposed mHealth tools and solutions) should pay special attention to those slices of the population where mobile smart phones are not used or unavailable for reasons of disadvantage, age, inequity, or other vulnerability. The burden is on policy makers, public health officials, data scientists and AI/ML developers to come together with affected stakeholders to figure out how to include these potentially left-out members of our communities in consequential policies, initiatives, and innovations. If anything, this crisis should be an opportunity to critically assess and redress elements of the digital divide that still define so much of contemporary society and that help to perpetuate more widespread societal inequities.

---

[12] Recently, scholars have been applying "fair" ML techniques directly to medicine. For example. Zink &Rose (2019) propose new "fair" ML modeling methods that use constrained & penalized regression to improve health insurance carrier risk adjustment for undercompensated groups; Pfohl, Marafino, et al. (2019) leverage adversarial learning and EHR data to develop a "fair" ASCD model; Pfohl, Duan, et al. (2019) uses counterfactual reasoning to apply fairness principles to clinical risk prediction at the individual level.

In this respect, applied concepts of fairness and health equity should not simply be treated *in the abstract* as self-edifying ideals or ornaments of justice that can be engineered into AI/ML technologies through technical retooling or interpolation. This approach will produce a blindered range of vision whereby only the patterns of bias and discrimination in underlying data distributions that can be measured, formalized, and statistically digested are treated as worthy and actionable indicators of inequity, and this to exclusion of the subcutaneous sociocultural dynamics of domination that slip through cracks of quantification (Fazelpour & Lipton, 2020). Rather, the existing sociohistorical, economic, and political patterns and qualities of disadvantage that create material conditions of injustice must be taken as the starting point for reflection on the impacts and prospects of technological interventions. This means that the *terminus ad quem* of any and all attempts to protect the interests of the vulnerable through the mobilization of AI/ML innovation should be anchored in reflection on the concrete, bottom-up *circumstances of justice*, in its *historical and material preconditions*. From this more pragmatic point of view (Dielman et al., 2017), there must be a prioritization of the real-world problems at the roots of lived injustice—problems that can then be treated as challenges "remediable" (Sen, 2011) by concerted social efforts and struggles for rectification, redistribution, and recognition (Fraser, 2010; Fraser & Honneth, 2003; Honneth, 2012). Only then will true-to-life demands for health equity and social justice be properly re-envisionable with and though the eyes of the oppressed. Only then will such demands become properly visible as struggles against the moral injuries inflicted by unjust social arrangements that obstruct the *participatory parity* of citizens in pursuing their unique paths to flourishing and in fully contributing to the moral and political life of the community.

# 5 Conclusion: Mobilizing responsible AI innovation to help today and to shape the society of tomorrow

The ethical challenges faced by those innovators, who are engaged in the second-front battle against COVID-19 have both immediate and intergenerational stakes. By carrying out their research and innovation ethically, transparently, and accountably, they will be better able to gain public trust, to accelerate collaborative problem-solving amidst a global community of scientists, to support the evidence-based clinical judgments of overtaxed doctors, to ease the immense and growing socio-economic hardships bore by most of present humanity, and to better prepare us for future pandemics.

But, these same innovators are confronted with dynamics of power and societal ills that together create conditions ripe for the abuse and misuse of the technological tools that they build and deploy. The data science and AI/ML community must therefore also act reflectively to safeguard cherished freedoms and values, the losses of which will very likely devastate our species for many generations to come. Taking this sort of anticipatory action is, however, well within its powers. Deliberate choices made, here and now, to engage in ethically informed and democratically governed innovation will not only help contemporary society build critical resistance to incipient strains of digital domination, it will facilitate the development of a future society that is more humane, rational, and enlightened.

For hundreds of years, at least since the 17th century dawning of the Baconian and Newtonian revolutions in the natural sciences, the drive to improve the human lot through the fruits of scientific discovery has guided the steady, albeit imperfect, forward progress of socially responsible innovation. Led by the torchlight of social conscience and reason, this collaborative project has relied on inclusive, equitable, and democratic practices of research that have simultaneously served as a model for the participatory attainment of legitimate social arrangements and therefore for the freedom and openness of modern society itself. Being true to their practices, responsible scientific researchers must now draw on these progressive energies to help steward humankind through this troubled time and into a better, more empowering, and more just species life for the society of tomorrow.

## Appendix: The Normative Dimension of Modern Scientific Advancement

Throughout this paper, I have eluded to the normative dimension of the history of modern scientific advancement. Though a full elaboration of this is beyond the scope of the current endeavour, the topic is worthy of some brief clarification and expansion. By making explicit the moral grammar underlying the practical success of modern scientific methods, we can begin to better discern a path toward the realisation of its beneficial potentials, while developing sightlines that will help us to steer clear of its greatest dangers. From this normative-historical perspective, the story of modern science is a story about how the successful development of a particular set of inclusive and consensus-based social practices of rational problem-solving carried out in the face of insuperable contingency has relied upon a corresponding release of the moral-practical potentials for cognitive humility, mutual responsibility, egalitarian reciprocity, individual autonomy, and unbounded social solidarity.[13]

As the broad-stroked narrative goes, at the very beginning of modernity, the deterioration of the religious and teleological order of things that typified traditional, pre-modern ways of life spurred the development of a thoroughgoing but salutary scepticism among a new generation of early modern scientists. The novel pressure to cope with the hardships of contingent reality without recourse to the authority of divine commandments or laws fixed by an intrinsically meaningful cosmic order (Taylor, 1989) consequently fuelled an increasing awareness of the inescapable uncertainty that seemed to define the epistemic fragility, fallibility, and finitude of the human condition (Blumenberg, 1983). Such a starting point in a reflexive acknowledgement of self-limitation and "learned ignorance" came to form the practical and epistemological basis of the experimental method of modern science (as initially exemplified in the work of pioneers like Pierre Gassendi, Francis Bacon, John Locke, and Isaac Newton).[14] This meant a shift from traditional modes of reasoning that appealed to the "inner nature of things and their necessary causes" to a new "science of experience" (Gassendi, 1624/1966) that was anchored in open-ended, collaborative problem-solving and carried out through ever-provisional forms of experimentation, reason-giving, and consensus-formation. In the midst of such a dramatic sociocultural sea-change, modern scientists became responsible to each other for sharing experience through standardized procedural mechanisms of rational warrant (like inductive reasoning and the experimental method) and for creating and reproducing the commonly-held vocabularies that alone could shape the possibilities of their innovation and discovery.

---

[13] Note that, in this Appendix, I am focussing on the constructive, normative dimension of the history of modern science and, for this reason, am leaving aside the empirical aspects of discursive and institutional power, politics, culture, and socio-economic stratification that inform other (equally important) critical-sociological histories of modern science (for instance: Foucault, 1961/1988, 1966/2007; Latour, 1993; Mirowski, 2002, 2011; Schaffer & Shapin, 1985; Shapin, 1996; 2010). While both normative and critical-sociological perspectives are crucial, I would suggest that it is also vital to resist disentangling them entirely. That is, from the perspective of a critical theory of society, one must endeavour to discover the sources of normativity that inhere pre-reflectively in concrete social and historical practices *per se*, and, only in this way, can one then gain the critical leverage needed to discern those distortions and malformations that manifest in both subtle and explicit forms of power, domination, and coercion (Honneth, 1993, 1995, 2009) . From this critical and ethical-practical point of view, one can gain access to historically-effective normativity by reconstructing the conditions of possibility of the particular social practices that lie behind human advancement. This does not mean vacating the interrogation of the sociohistorical fields wherein dispersed and concentrated patterns of violence and power inhere but rather taking a performatively consistent approach to the critique of the latter by first clarifying and making explicit what has gone moral-practically awry. It is this last bit that I'm concentrated on here.

[14] Admittedly, the generic idea of a "scientific method" is overly schematic as has been pointed out by Medawar (1967), Shapin (2007), and others. Likewise, reference to a unifying idea of "modern science" has been usefully deconstructed by historians and philosophers of science (for example in Cartwright, 1999; Dupre, 1993; Galison & Stump, 1996). I use these terms as sign posts for particular kinds of distinctively post-conventional social practices that carry historically sustained normative relevance rather than as descriptors that are applicable in their historical specificity.

Thinkers from Charles Sanders Peirce and John Dewey to Karl-Otto Apel, Robert Brandom, and Juergen Habermas have long emphasized the importance of reconstructing the normative presuppositions that lie behind the emergence of these consensus-based and procedurally rational social practices. On this view, grasping the moral-practical enabling conditions for this new way of concerted human coping can help us to better comprehend the social and ethical determinants that have weighed heavily in the success of modern science itself. There are several.

The first is ***the necessity of cognitive and methodological humility***. Dewey, in this connection, points out the "importance of uncertainty" (Dewey, 1933/1997, p. 12) and of the unending need to draw upon what Peirce (1877) called the "irritation of doubt" (p. 233) in the pursuit of open scientific inquiry, tentative suggestion, and experimentation. A starting point in cognitive and methodological humility functions as a lynchpin of the essential corrigibility and incompletability of modern scientific research and innovation. It secures the "unprejudiced openness that characterizes [its] cognitive process" (Habermas, 1992, p. 36). That no fallible interlocutor is entitled to have the last word in matters of scientific investigation leaves each participant in the unbounded community of inquiry (Apel, 1998; Peirce, 1868) no choice but to speak, to ask of others "Why?," to demand from them reasons for their claims and conclusions that are continuously liable to a mobile tribunal of ongoing rational assessment, criticism, and further observation (Leslie, 2016). The indeterminate and anti-authoritarian character of this open process of modern scientific inquiry unlocks trajectories of indefinite improvement at the same time as it lines up with the "inescapable incompleteness" (Rogers, 2009, p. xii) of modern democratic ways of life, which are legitimated by persistent practices of discursive exchange and meaning redemption. Directly drawing inspiration from "the spirit and method of science," Dewey summarizes, "the prime condition of a democratically organized public is a kind of knowledge and insight which does not yet exist… An obvious requirement is freedom of social inquiry and of distribution of its conclusions" (Dewey, 1927, p.166).

Already implicit in the practical concomitants of the demand for methodological humility is a second normative precondition for the success of modern science: ***the imperative of publicity and the responsibilities of communication and listening***. An unbounded community of scientific inquiry can endure as such only insofar as it is organized around "free and systematic communication" (Dewey, 1927, p.167). The tentative and corrigible character of modern scientific insights makes this kind of publicity necessary inasmuch as inclusive debate and conversation are needed to ensure the continuous revision of beliefs and to foster the enlargement of an evolving space of scientific creativity and innovation (Mill, 1859/2006). This entails that "no one who could make a relevant contribution concerning a controversial validity claim must be excluded" (Habermas, 2008, p. 50). Likewise, all relevant positions, opinions, and information must be aired, exchanged, and weighed so that the stance participants take can be motivated "by the revisionary power of free-floating reasons" (Ibid.). The boundless conversation that underwrites the advancement of scientific inquiry must, along these lines, be open and accessible to all. Scientists have a responsibility to communicate their ideas plainly and to as wide an audience as possible, and non-scientist members of the public have a corollary responsibility to listen (Asimov, 1987; Leslie 2020).

A third normative precondition for the advancement of modern science stems from the ***normative standing of participants involved in the ongoing rational dialogue of the scientific "communication community"*** (Apel, 1998, p. 225). In order for those engaged in practices of giving and asking for reasons *to be conferred the authority* to endorse validity claims and, in turn, *to be held accountable* for their commitments to these, they must reciprocally grant each other normative status as being rational and responsible agents (Brandom, 1994/2001, 2000, 2013). The mutual conferral of this normative standing operates as a pragmatic presupposition of communicative practices of scientific inquiry, for it makes interlocutors liable to each other for the rational assessment of the arguments they tender. Beyond this, the process of rational assessment itself entails a further set of procedural requirements that place additional normative-pragmatic demands on participants engaged in inquiry. Because the claims to propositional truth that are building blocks of modern science carry an unconditional, context-bursting force that reaches beyond the embodied and factually-situated circumstances in which they are uttered, these claims structurally mandate procedures that, at once, "guarantee the impartiality of the process of judging" (Habermas, 1990/2001, p.122) and secure an "egalitarian universalism" (Habermas, 2008, p.49) in the practices of giving and asking for reasons by which propositions gains rational acceptability. Chief among such unavoidable idealizing suppositions of those engaged rational discourse are mutual respect, egalitarian reciprocity, equal right to engage in communication and equal opportunity to contribute to it, non-coercion, participatory parity, and sincerity (Habermas, 1990/2001, 1992, 1996, 1998, 2008).

A final normative precondition, ***the intrinsic sociality of science***, is predicated on the role that scientific inquiry plays as a practical medium of problem-solving through collaboration, reason-giving, and experimentation. Although the explanatory ambitions of the modern natural sciences have largely been anchored in making truth claims about the world through physical observations, quantified measurements, and the experimental practices of hypothesis testing, these approaches are, at bottom, rooted in social processes of intersubjective communication that are driven by shared endeavours to cope with challenges deemed worthy of response. Scientific practices are always already embedded in a community of interpretation and in holistic contexts of individual life plans and collective social projects (Apel, 1998; Royce, 1908/1995;). The evolution of scientific inquiry occurs within a changing space of reasons, interpretations, and values (Apel, 1984, 1999; Sellars, 1956/1997; Taylor, 1964/1980; Von Wright, 1971/2004). And, as humans adjust their purposes and goals to meet the needs of their times, science too changes its focus, outlook, and direction. This holistic and value-oriented departure point of scientific practices implies that modern science should not be viewed, first and foremost, as operationally independent from human beliefs, aims, and interpretations, but rather as an ethically implicated set of problem-solving practices that are steered by the values and commitments of its embodied producers. In Dewey's words, "The notion of the complete separation of science from the social environment is a fallacy which encourages irresponsibility, on the part of scientists, regarding the social consequences of their work" (Dewey, 1938, p. 489). This intrinsic sociality of science functions then as an enabling condition of the responsibility of innovation and of its humane pursuit of what Francis Bacon called the "relief man's estate [through discovery]" (Bacon, 1605/2001).

To close here, it may be useful to note that, taken together, these normative-pragmatic presuppositions of the advancement of modern science continuously push researchers and innovators to think beyond themselves and their existing communities of practice to consider their role in safeguarding the endurance of a greater living whole. Prompted to see themselves in this light, they are better equipped to embrace the essential positions they occupy both as stakeholders vested in a world-yet-to-come and as committed members of two broader, expanding circles. Their participation in the first of these involves playing an active albeit transient part in an unbounded community of learning and discovery that is charged with advancing the "permanent interests of humankind as a progressive being," to paraphrase J.S. Mill (1859/2006). The execution of such a species-level commission to improve the present and future conditions of life demands that, not only scientists, but all members of humanity be able to carry out the indefinite and transgenerational tasks of shared knowledge-creation and collaborative world-making through unfettered communication, consensus-based value articulation, and deliberative will formation. To this end, humankind itself must become ever more capable of inclusively cultivating and drawing upon the unique talents, passions, and callings of each of its increasing number. That is, as this circle of shared learning and discovery expands, every human being should be capacitated to pursue their own path to intellectual and creative self-realisation so that the universal fulfilment of the full potential of each can usher forward the greater social project of the sustenance and flourishing of all. This civilizational impetus to "fully integrated personality" (Dewey, 1946, p. 148) entails that any arbitrary socioeconomic or geopolitical barriers to equitable flourishing be demolished so that no future Ramanujan, Curie, Turing, or Einstein can be lost to the dynamics of societal oppression that stamp out the flames of human genius before they have a chance to ignite.

The second expanding circle in which researchers and technologist are included extends the responsibility of innovation to all members of the circle of life itself. Outfitted with multiplying technological capacities to bring about species self-annihilation, mass extinction, and biospheric catastrophe, the human community of learning and discovery now finds itself implicated as a potentially cataclysmic *force of nature*, in its own right. Those at the tiller of scientific research and innovation are consequently no longer entitled to simply assume a kind of legitimate epistemological or ontological division between "nature" and "society." The presumption of such a "great divide" between natural and cultural worlds has promoted a misdirected self-perception among scientists that they are engaged in a neutral and value-free enterprise, thereby enabling reckless strains of the modern natural sciences to claim functional immunity from the curbing modes of ethical critique that derive from social environments in which they are situated. It has also allowed them to instrumentally treat the living and inanimate constituents of the natural world simply as objects available for appropriation, calculation, and control. With the ushering in of the Anthropocene epoch such a presumed dichotomy between nature and society has become increasingly implausible inasmuch as the scope of the anthropogenic impacts on climate and biosphere increasingly inculpates humankind not only as a natural force of geohistorical consequence but as an essential co-originator of the conditions of possibility for the survival and flourishing of life on earth. In this way, nature *as such* can no longer be seen merely as an object to be measured and manipulated but is now *as us*, above all, implicated as a subject bound by ethical obligations of existential import and intergenerational reach.

It follows from all this that the human community is part and parcel of a wider circle of natural organisms whence, over a fortunate 3.7-billion-year trajectory of evolutionary transformation, it has developed the exceptional capabilities for limitless creation and mass destruction for which it is, in the end, uniquely accountable. We should take cognizance here, however, that, although modern science has been an essential catalyst in facilitating the enabling conditions of these dangerous competences, it has concurrently shown a path to the societal acceptance of such a unique responsibility by spurring an ethical self-understanding of humanity's place in exactly this deep history of life. From Hutton and Lyell to Sedgwick and Darwin, modern scientific insights have enabled chastening and worldview-decentering access to a widening temporal frame of geological and evolutionary history within which the human species has been placed on a living continuum extending from the very first unicellular organisms to the shrinking plurality of flora and fauna that typifies our current era of biodiversity drain and the humanly prompted "sixth extinction". From the sharp end of the deep historical arc at which we now find ourselves, it is possible to peer back across thousands of millennia so to see that all biological individuals have been interlinked in such an evolving circle of life from the outset and that, notwithstanding the contemporary anthropogenic mass extermination of species, living matter's diversifying impetus has tracked the development of a kind a holistic unity within the unbounded community of the biospheric whole.

From the vista of humankind's membership in this broader biotic totality, our story may well be seen as a tale of two species. For, on the one hand, we are a species principally unconstrained in its pursuit of the boundless possibilities opened by the infinite generativity of its capacity for language, representation, and symbolic experience—a species which is, for precisely that reason, readily capacitated to effect the self-annihilation of planetary life—a dangerous species. On the other, we are a species endowed with recourse to the media of collaboration, communication, and criticism by means of which we are able to constrain our penchant for technoscientific hubris and to sustain the futurity and flourishing of the greater living whole. We are an ethical species endowed with a deep historically ingrained sense of responsibility to the intrinsic worth of life as such and hence capable of stewarding the sustenance of the biosphere as its trustees and as its guardians. It is perhaps the greatest redeeming power of modern scientific advancement that it has granted us the wherewithal to tell this second, moral story.

## Acknowledgements

Not so many moons ago, this paper began as a short blog, written for the Alan Turing Institute, as the UK entered into COVID-19 lockdown. That initial piece quickly grew tight in its own digital skin. As it developed into a longer and more academic article, which aimed to reach a general audience but remain pertinent for specialists, thoughtful and incisive input from four anonymous reviewers and the editor-in-chief at the HDSR stewarded many improvements, though the deficiencies that remain in it are its author's responsibility alone. I would like to thank them as well as my research assistant, Christina Hitrova, for her efforts in helping me get together the bibliography. Corianna Moffatt's assistance with copyediting and formatting also proved invaluable as did helpful comments from Christopher Burr and Suzanne Smith, who both worked through the paper as it neared its completion.